\documentclass[fleqn,usenatbib]{mnras}

\usepackage{newtxtext,newtxmath}

\usepackage[T1]{fontenc}
\usepackage{ae,aecompl}

\usepackage{graphicx}	
\usepackage{amsmath}	

\usepackage{amssymb}	
\usepackage{caption}
\usepackage{subcaption}
\usepackage{booktabs}
\usepackage{longtable}
\graphicspath{{Figures/}} 

\newcommand{\ergps}{erg s\textsuperscript{-1}} 
\newcommand{\Nfof}{N_\text{fof}}

\defcitealias{Mantz2010TheRelations}{M10}
\newcommand{\Mantz}{\citetalias{Mantz2010TheRelations}}
\defcitealias{2015AA...573A.118Lovisari_Scaling}{L15}
\newcommand{\Lovisari}{\citetalias{2015AA...573A.118Lovisari_Scaling}}
\defcitealias{2015MNRAS.449.3806Anderson_scaling}{A15}
\newcommand{\Anderson}{\citetalias{2015MNRAS.449.3806Anderson_scaling}}
\defcitealias{Andreon2017}{A17}
\newcommand{\Andreon}{\citetalias{Andreon2017}}
\defcitealias{Giles2017ChandraRelation}{G17}
\newcommand{\Giles}{\citetalias{Giles2017ChandraRelation}}
\defcitealias{Chiu2022TheSurvey}{C22}
\newcommand{\Chiu}{\citetalias{Chiu2022TheSurvey}}

\defcitealias{2018XXL_XX_365GC}{XXL Paper XX}
\newcommand{\PaperXX}{\citetalias{2018XXL_XX_365GC}}
\defcitealias{2018AA...620A..12C_3XLSS}{XXL Paper XXVII}
\newcommand{\PaperXXVII}{\citetalias{2018AA...620A..12C_3XLSS}}
\defcitealias{Crossett2022TheEmission}{XXL Paper XLV}
\newcommand{\PaperXLV}{\citetalias{Crossett2022TheEmission}}
\defcitealias{XXLXXIV-Faccioli18}{XXL Paper XXIV}
\newcommand{\PaperXXIV}{\citetalias{XXLXXIV-Faccioli18}}
\defcitealias{XXLIII}{XXL Paper III}
\newcommand{\PaperIII}{\citetalias{XXLIII}}
\defcitealias{XXL1}{XXL Paper I}
\newcommand{\PaperI}{\citetalias{XXL1}}
\defcitealias{XXLII}{XXL Paper II}
\newcommand{\PaperII}{\citetalias{XXLII}}
\defcitealias{Giles2022TheSurveys}{XXL Paper XLII}
\newcommand{\PaperXLII}{\citetalias{Giles2022TheSurveys}}

\title[XXL: X-ray properties of optically Selected Groups]{\centering The XXL Survey\thanks{Based on observations obtained with XMM-Newton, an ESA science mission with instruments and contributions directly funded by ESA Member States and NASA} \\
LIV. X-ray Luminosity Function and Luminosity-Mass Relation of Optically Selected Galaxy Groups}

\author[C. Wood et al.]{ 
C. Wood,$^{1}$\thanks{E-mail: cai.wood@bristol.ac.uk}
B. J. Maughan$^{1}$,
J. P. Crossett$^{2}$, 
D. Eckert$^{3}$, 
M. Pierre$^{4}$, 
M. E. Ramos-Ceja$^{5}$,
\newauthor
A. S. G. Robotham$^{6}$,
C. Adami$^{7}$,
L. Faccioli$^{4}$,
E. Koulouridis$^{8}$, 
S. L. McGee$^{9}$,
F. Pacaud$^{10}$
\newauthor
and S. Phillipps$^{1}$
\\
$^{1}$H.H.Wills Physics Laboratory, University of Bristol, Tyndall Avenue, Bristol, BS8 1TL, UK\\
$^{2}$Instituto de Física y Astronomía, Universidad de Valparaíso, Avda. Gran Bretaña 1111 Valparaíso, Chile\\
$^{3}$Department of Astronomy, University of Geneva, ch. d’Écogia 16, CH-1290 Versoix, Switzerland\\
$^{4}$Université Paris-Saclay, Université Paris Cité, CEA, CNRS, AIM, F-91191, Gif-sur-Yvette, France\\
$^{5}$Max-Planck Institut für extraterrestrische Physik, Postfach 1312, 85741 Garching bei München, Germany\\
$^{6}$ICRAR, University of Western Australia, Crawley, WA 6009, Australia\\
$^{7}$Aix Marseille Univ., CNRS, CNES, LAM, Marseille, F-13007, France\\
$^{8}$IAASARS, National Observatory of Athens, GR-15236 Penteli, Greece\\
$^{9}$School of Physics and Astronomy, University of Birmingham, Birmingham, B15 2TT, UK\\
$^{10}$Argelander Institut f\"ur Astronomie, Universit\"at Bonn, D-53121 Bonn, Germany
}

\date{Accepted XXX. Received YYY; in original form ZZZ}

\pubyear{2025}

\begin{document}
\label{firstpage}
\pagerange{\pageref{firstpage}--\pageref{lastpage}}
\maketitle


\begin{abstract}

The overlap between the GAMA spectroscopic survey and the XXL X-ray survey was used to study the X-ray properties of optically-selected groups of galaxies. 
Forced X-ray aperture photometry was applied to an optically-selected sample of 235 groups (containing at least five member galaxies) to measure their X-ray luminosities in the regime of low signal to noise X-ray data. 
The sample encompasses X-ray luminosities over an order of magnitude fainter than typical X-ray selected samples, and avoids X-ray selection biases. 
This gives access to low mass groups where the effects of non-gravitational processes, such as AGN-feedback, should be most apparent and could inhibit their detection in an X-ray survey. 
We measured the X-ray luminosity function (XLF) of the sample, and found it to be consistent with the extrapolation of the XLF from X-ray selected samples at higher luminosities. 
The XLF was combined with a theoretical halo mass function to infer the form of the scaling relation between X-ray luminosity and mass (LM relation) for the GAMA groups. 
We found a slope of $1.87 \pm 0.12$, which is steeper than self similarity in this mass regime. 
When comparing with other measurements of the LM relation, we find evidence for a steepening of the slope in the low mass regime, likely due to the impact of non-gravitational processes. 
Our approach can be translated to eROSITA data using multi-wavelength surveys to constrain the X-ray properties of galaxy groups in the limits of high redshift and low mass.

\end{abstract}




\section{Introduction}
\label{sec:Intro}

As the largest gravitationally bound structures in the Universe, galaxy clusters are interesting astrophysical laboratories and important cosmological probes \citep[e.g.,][]{2011ARAA..49..409Allen_ObsCLusters, Vikhlinin2014ClustersGalaxies, 2019SSRv..215...25Pratt_ClusterReview}. 
Galaxy clusters comprise tens to thousands of galaxies plus an X-ray emitting intra-cluster medium (ICM) bound by a dark-matter-dominated gravitational potential well.
While the investigation of high-mass galaxy clusters has a rich history, the lower-mass galaxy groups are intrinsically fainter and hence harder to find and study.
As there is no definitive boundary in mass between groups and clusters, a useful working definition is $\sim 10^{14} M_{\odot}$ \citep[e.g.,][]{2013SSRv..177..247Giodini_LMscale, Kolcu2022QuantifyingGroups}, and in this paper the term "groups" will be used to collectively refer to groups and clusters.

In order to determine the characteristics and abundances of galaxy groups it is necessary to catalogue statistically complete samples. 
The results obtained from analysing such samples are used to determine number density functions and scaling relation laws for various observable properties.
These can then be compared to theoretical models in order to constrain the details of astrophysical processes and the values of cosmological parameters.

The two original galaxy cluster catalogues are those of \cite{1958ApJS....3..211Abell} and \cite{1961cgcg.book.....Z}, and both were constructed by counting over-densities of galaxies in optical images.
The earliest reasonably complete catalogue \cite{1975gaun.book..557D} was reconstructed using a quantitative method to create the first reproducible cluster catalogue based on galactic positions and redshifts \citep{1982ApJ...257..423HuchraGellerGalCat}.
More recently, the fact that galaxies located in the cores of galaxy clusters have older stellar populations, and so form a tight red-sequence in colour-magnitude space, led \cite{Gladders2000AAlgorithm} to create a red-sequence based cluster finding algorithm which was applied in the construction of the first Red-sequence Cluster Survey \citep{2005ApJS..157....1GladdersRedClusterSurvey}.
\cite{2007ApJ...660..239Koestar_MaxBCG_cat} and \cite{2007ApJ...671..153YandSDSS} applied this method to the Sloan Digital Sky Survey \citep[SDSS,][]{Abazajian2004TheSurvey} to detect galaxy clusters based on their brightest cluster galaxy and location in position and redshift space. 
These approaches are being used to detect clusters out to high redshifts in current and upcoming optical and IR surveys \citep{Ivezic2019LSST:Products, Amendola2018CosmologySatellite}.
Galaxy redshift surveys are also used to find galaxy clusters by applying a clustering or friends-of-friends (FoF) algorithm \citep[e.g.,][]{Eke2004GalaxyCatalogue, Yang2005A2dFGRS, GAMA}.
These reduce the impact of projection effects in photometric surveys, but at greater observational cost.

While the largest cluster catalogues to date have been constructed using optical data, X-ray surveys yield robust cluster samples since the presence of an X-ray bright ICM is an unambiguous sign of a deep gravitational potential.
This leads to the most statistically well defined cluster samples. 
The \textit{ROSAT} All-Sky Survey \citep[RASS,][]{1993Sci...260.1769TrumperRASS} provided the foundation for large, high-quality X-ray selected samples of galaxy clusters, such as BCS \citep{Ebeling1998TheDistribution}, REFLEX \citep{2004AA...425..367Bohringer_REFLEXI}, and REFLEX II \citep{REFLEXIV}.
\textit{ROSAT} was also used for many serendipitous X-ray cluster surveys such as the 400 square degree survey \citep{1998ApJ...502..558Vikhlinin_160SD, 2007ApJS..172..561Burenin_400SD} and the Wide Angle \textit{ROSAT} Pointed Survey \citep[WARPS,][]{2008ApJS..176..374Horner_WARPS}.
\textit{XMM-Newton} and \textit{Chandra} have also been used for serendipitous cluster detections such as the XMM Cluster Survey \citep{2012MNRAS.423.1024Mehrtens_XCS}, the X-Class survey \citep{Koulouridis2021The1.5} and the \textit{Chandra} Multiwavelength Project \citep{2006ApJ...645..955Barkhouse_2006_CHaMP}, along with targeted cluster surveys such as XXL \citep[hereafter \PaperI]{XXL1}.
The Spectrum-Roentgen-Gamma mission carrying the extended Roentgen Survey with an Imaging Telescope Array \citep[\textit{eROSITA},][]{Predehl2021TheSRG} aims to create an all-sky X-ray survey which is 25 times more sensitive than the RASS in the soft X-ray band.
Upon completion, the \textit{eROSITA} all-sky survey is expected to have detected of order 100 thousand rich clusters of galaxies \citep{Merloni2012EROSITAUniverse, Predehl2021TheSRG}, which would be about two orders of magnitude larger than current X-ray cluster samples. 

Cluster surveys are also carried out in the millimetre using the Sunyaev-Zel'dovich (SZ) effect, such as the South Pole Telescope (SPT) Survey \citep{2013ApJ...763..127R_SouthPoleTelSurvey}, the all-sky \textit{Planck} \citep{Ade2016PlanckSources}, and the Atacama Cosmology Telescope \citep{Hilton2021TheClusters}.
Due to the redshift independent nature of the SZ signal, such surveys are an increasingly important tools for cosmological studies with clusters.
SZ surveys have thus far been limited to relatively high mass clusters due to the sensitivity limitations of the survey instruments, but sensitivity is constantly improving.
For example, the SPT-3G receiver on the SPT is expected to detect every group with $M_{500c}$\footnote{$M_{500c}$ is the mass within spherical volume of radius $R_{500c}$ with a density 500 times the critical density at that redshift}$ > 10^{14} M_\odot$ \citep{Benson2014SPT-3G:Telescope, Sobrin2022TheSPT-3G}.

Assuming galaxy groups form solely through gravitational mergers, following the hierarchical formation model, it is expected that all clusters should be scaled versions of each other provided the change in density of the Universe is considered.
The assumption of a self-similar model \citep{1986MNRAS.222..323Kaiser_SelfSimilar} predicts power-law scaling relations between cluster properties such as luminosity, temperature and mass.
Previous work on scaling relations has typically been confined to the study of relatively massive systems, from X-ray selected samples, and using per-cluster mass estimates to allow the direct modelling of scaling relations.
Examples of mass measurement techniques include hydrostatic masses \citep{Vikhlinin2006Relation}, weak lensing masses \citep{Mellier1999ProbingLensing}, and caustic masses \citep{Diaferio1997InfallClusters}.
When measured observationally, these scaling relations are generally found to differ from self similar expectations. 
This is most prominent for low mass systems, while some studies of high mass clusters do find self-similar behaviour in some scaling relations \citep[e.g.,][]{Maughan2012Self-similarRelation, Mulroy2019LoCuSS:Content}.
The departures from self-similarity are most noticeable for the relations which are dependant on density and distribution of ICM (e.g. luminosity, gas mass).
These properties are found to show steeper correlation with mass and temperature than expected \citep{2007ApJ...668..772Maughan_LM,  2013SSRv..177..247Giodini_LMscale,  2016MNRAS.461.3222DeMartino_LMscaling, Lovisari2021ScalingGroups}.
This behaviour is broadly understood to be a result of non-gravitational processes that are ignored by the self-similar model.
These processes (e.g., active galactic nuclei (AGN) feedback, cooling, star formation, supernovae) are expected to disproportionately affect the properties of the ICM in lower mass systems where the gravitational potential well is shallower.
This makes galaxy groups a useful experimental tool for studying these feedback processes.
For detailed reviews of the scaling properties and AGN feedback in galaxy groups see \cite{Lovisari2021ScalingGroups, Eckert2021FeedbackGroups}.

The impact of feedback may also be apparent in the X-ray luminosity function (XLF).
This measure of the number density of groups per luminosity interval can be calculated straightforwardly from limited survey data, where mass and temperature estimates are not well constrained. 
The XLF allows for the study of the development of structure and evolution of baryonic properties in the Universe \citep{1979ApJ...232..689Bahcall_XLF, 1982ApJ...253..485Piccinotti_XLF, 1998ApJ...498L..21Vikhlinin_XLF, 2003MNRAS.342..287Allen_XLF}.
\cite{2004ApJ...607..175Mullis_XLF} found significant evidence for negative evolution of the XLF at the bright end, implying that the number density of high luminosity groups was lower at high redshifts than in the local Universe.
This negative evolution was also observed by \cite{2013MNRAS.435.3231Koens_WARPS} and \citet[hereafter \PaperXX]{2018XXL_XX_365GC}.
The XLF for the local Universe has also been well studied using \textit{ROSAT}, and when combined with a scaling relation between X-ray luminosity and mass, may be used to infer the cluster mass function and hence constrain cosmological parameters \citep{2008MNRAS.387.1179Mantz_XLF, REFLEXIV}.
\citet{Liu2022TheGroups} fits the XLF for the cluster sample detected in the \textit{eROSITA} Final-Depth Survey, and finds good agreement with the results obtained by other recent X-ray surveys.
Studies based on all-sky surveys give the best determination of the XLF for the local Universe, while serendipitous surveys extend our knowledge out to higher redshifts and fainter luminosities (i.e. lower masses).
A large limitation in the determination of the XLF is the detection limit of X-ray surveys resulting in the faint end being poorly constrained.

The aim of the present work is to measure the XLF of optically-selected groups in order to better constrain the low-luminosity end.
Using the spectroscopically-complete sample from the Galaxy And Mass Assembly survey \citep[GAMA,][]{2011MNRAS.413..971Driver_GAMA} which overlaps with the XXL X-ray survey (\PaperI), a statistically complete sample of optically selected galaxy groups can be obtained.
This optical selection allows the inclusion of groups that would be too faint to be included in an X-ray selected sample, and also ensures the sample is unbiased with respect to the X-ray properties (such as the gas fraction, X-ray morphology, or the presence of cool cores \citep{Eckert2011TheHIFLUGCS, Andreon2016TheSample, Rossetti2016MeasuringOffset}) in an approach somewhat similar to that taken by \citet{2011A&A...536A..37A}. 
The shape of the XLF for the GAMA selected sample will then be compared to previous studies of the XLF done using X-ray selected samples.

Following on from the XLF, this paper explores the shape of the X-ray luminosity - mass (LM) scaling relation in the low luminosity and hence low mass regime.
Measurements of the LM relation for lower mass groups are less common, due mainly to the difficulty in obtaining mass estimates for these fainter systems. 
Similarly, measurements of the LM relation for optically-selected samples, while desirable due to the minimisation of X-ray selection biases, are also rare due to the difficulties in obtaining uniform X-ray data and mass estimates.
Our analysis is thus somewhat unusual in the use of uniform X-ray data, the low masses of the systems, and the use of the XLF and the halo mass function to infer the LM relation. 
From self-similarity, we would expect the slope of the LM relation to be $4/3$ \citep{Maughan2006The1.0} for bolometric luminosities ($\approx 1$ for soft band luminosities), however, previous evidence shows steeper slopes than expected from self similarity.
The slope derived from self-similarity assumes emission only from thermal bremsstrahlung, however, in the galaxy group regime temperatures drop below 2 keV where line emission is prominent.
This would result in the luminosity from lower mass systems being higher than expected from self-similarity, thus flattening the slope of the LM relation.
Therefore, in the mass regime covered by this work, any steepening of the LM relation would be a lower limit of the impact of non-gravitational processes.

The X-ray properties of the GAMA selected groups in XXL have also been studied by \citet[hereafter \PaperXLII]{Giles2022TheSurveys} and \citet[hereafter \PaperXLV]{Crossett2022TheEmission}.
\PaperXLII{} examined the scaling relation between velocity dispersion and temperature for the groups detected by both GAMA and XXL, while \PaperXLV{} explores the correlations between X-ray and optical properties of the GAMA sample.
In the latter, forced X-ray aperture photometry on the GAMA group locations was performed based on the method by \citep{2018MNRAS.477.5517Willis}.
Due to similarities in approach to this paper, a brief comparison of the results from \PaperXLV{} will be discussed below.

This paper is organised as follows.
In section \ref{sec:GAMAXXL} we describe the GAMA and XXL sample selections.
The X-ray forced aperture photometry methods are described in Section \ref{sec:photometry}.
In section \ref{sec:XLF} we detail our methodology and present the XLF.
In section \ref{sec:LM}, we detail how we use the XLF to infer the shape of the LM relation and present our best fit parameters.
We discuss our findings, possible systematics, and possible implications for the impact of non-gravitational processes in Section \ref{sec:Discussion} and draw our conclusions in Section \ref{sec:conclusions}.
This work assumes the WMAP9 cosmology from \cite{Hinshaw13} with $H_0 = 69.32$, $\Omega_M$ = 0.2815, $\Omega_R$ = 0, $\Omega_\Lambda$ = 0.7185, and $\sigma_8 = 0.82$.

\section{GAMA and XXL Catalogues}
\label{sec:GAMAXXL}

The GAMA \citep{2011MNRAS.413..971Driver_GAMA} spectroscopic survey used the \textit{Anglo-Australian Telescope} to observe 300,000 galaxies in a 286 deg$^2$ region down to an r-band magnitude of $r<19.8$ with a spectroscopic completeness $>95\% $ \citep{Baldry2018Galaxy3}.
A GAMA galaxy group catalogue containing 26,194 galaxy groups was generated using a FoF algorithm \citep{GAMA}.
The grouping algorithm was calibrated on an extensive set of mock catalogues to optimise completeness, minimise contamination, and evaluate recovery of the underlying predicted halo population, and their properties, in the group mass regime.
Groups with five or more members were found to give a meaningful estimate of the dynamical velocity dispersion.
Our study therefore makes use of only those GAMA groups with five or more members, and this defines the selection criteria for the GAMA group sample.
This has a redshift range of $0.03 \leq z \leq 0.45$ and an approximate mass range of $11.1 \leq \log\text{M}_{200}/\text{M}_\odot \leq 15.5$, with only more massive groups detectable at the higher redshifts.
\citet{Driver2022AnData} finds that the GAMA group sample has a mass completeness limit of approximately $10^{12.8} M_{\odot}$, below which the sample gradually becomes irrecoverably incomplete.
For a detailed description of the GAMA group catalogue along with the FoF algorithm and its parameters, see \cite{GAMA}.

The XXL Survey is an X-ray survey using \textit{XMM-Newton} covering 50 deg$^2$ across two extragalactic fields (\PaperI) to a sensitivity of $6\times10^{-15}$ erg s$^{-1}$ cm$^{-2}$ (for point-sources) in the [0.5-2] keV energy band. 
The key science goal of XXL is to create a well-defined X-ray selected sample of galaxy groups suitable for cosmological studies.
The XXL group catalogue (\PaperXX) contains a complete subset of groups for which the selection function is well determined, plus all X-ray groups which were spectroscopically confirmed before the release date.
In addition to a group catalogue, XXL released a point source catalogue \citet[hereafter \PaperXXVII]{2018AA...620A..12C_3XLSS} and the analysis pipeline used by XXL for both catalogues is detailed in \citet[\PaperXXIV]{XXLXXIV-Faccioli18}.
The XXL catalogues used in this paper were processed using version 3.3 of the \texttt{Xamin} pipeline.
The XXL group catalogue is primarily made up of uncontaminated (C1) groups, selected based on apparent size and the likelihood of a source being extended.
C1 groups are complemented by a second, typically fainter, sample (C2) which have a lower purity (estimated to be $\sim50\%$), and spectroscopically-confirmed C2 groups are included in the XXL group catalogue.

\begin{figure}
    \includegraphics[width=\columnwidth]{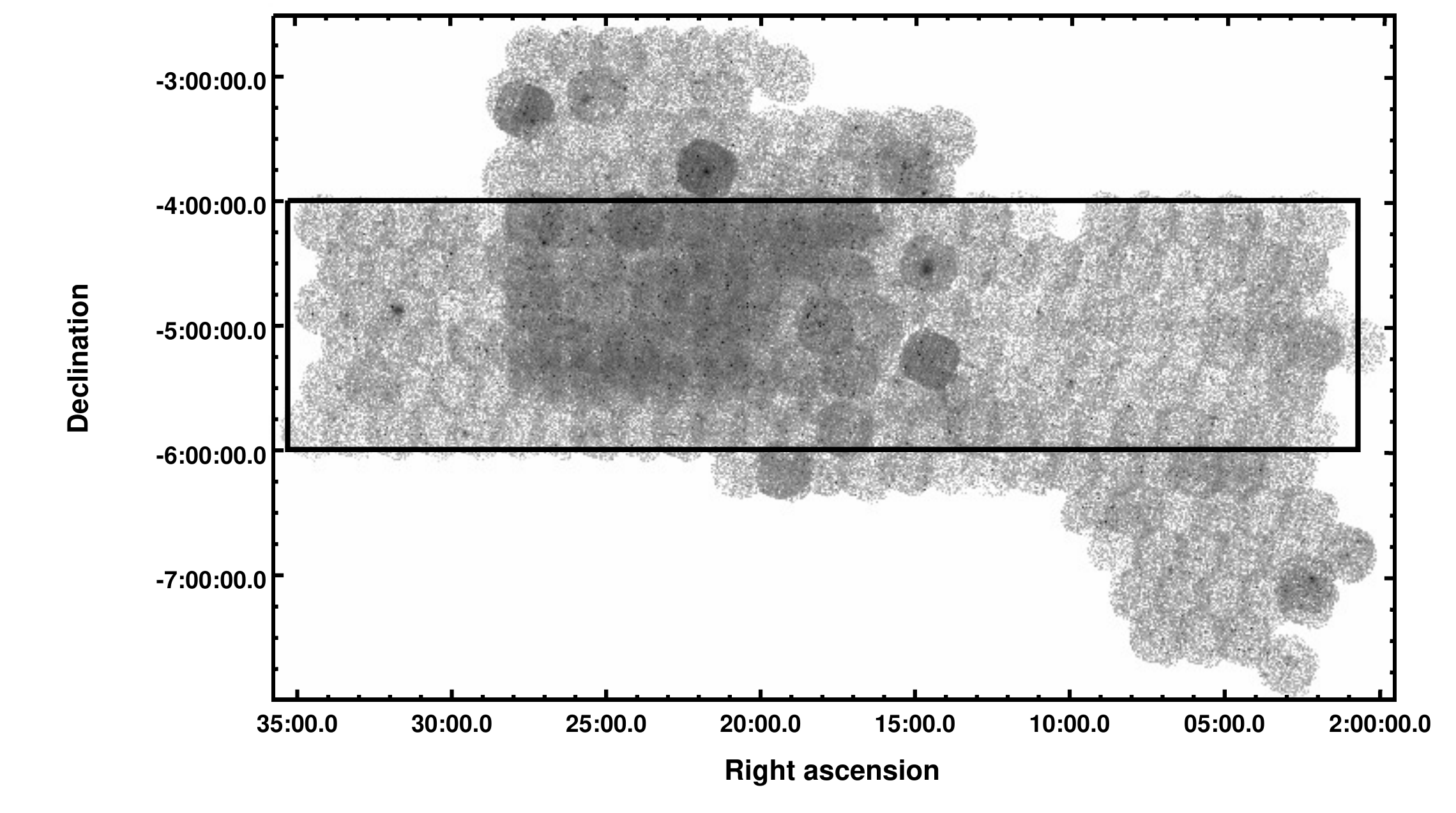}
    \caption{XMM-Newton image of the XXL North field in the $0.5-2$ keV band, with the box highlighting the overlapping GAMA region. 
    }
    \label{fig:Overlap_Region}
\end{figure}

In our analysis, we perform X-ray photometry to estimate the X-ray luminosity of the GAMA groups. 
For this we use a mosaic of \textit{XMM-Newton} pointings of the XXL North field in the $0.5-2$ keV band, along with corresponding exposure maps.
The mosaics were produced by combining the count maps and summing up the pixels for the three European Photon Imaging Camera (EPIC) instruments aboard \textit{XMM-Newton}, which consists of one PN (positive/negative) detector and two EPIC Metal Oxide Semi-conductor (MOS) detectors.
The exposure maps were weighted and summed to produce an exposure map at the equivalent depth of a single MOS observation.
The 16.6 $\text{deg}^2$ overlap region between the GAMA and XXL surveys is shown in Figure \ref{fig:Overlap_Region}.
The overlap region contains 235 GAMA galaxy groups with five or more members, with statistics for the region detailed in Table \ref{tab:NumberofGroups}.

\begin{table}
    \centering
    \caption[Summary of numbers of GAMA groups]{The numbers of galaxy groups in different subsets in the GAMA/XXL overlap region and limited by the maximum redshift of the G02 GAMA sample ($z \leq 0.46$). Here we consider GAMA groups with $\geq 5$ members, and XXL groups from the \PaperXX{} catalogue.}
    \label{tab:NumberofGroups}
    \begin{tabular}{lc}
         Description & Number \\
         \hline
         GAMA groups & 235 \\
         XXL groups & 94 \\ 
         GAMA groups with XXL group matched & 57 \\
         GAMA groups with X-ray point source within 30\" & 25 \\
         GAMA groups with X-ray point source within 90\" & 96 \\
         GAMA groups with X-ray point source within 110\" & 117 \\ 
    \end{tabular}
\end{table}

\section{X-ray photometry of GAMA groups}
\label{sec:photometry}

\begin{figure}
    \includegraphics[width=\columnwidth]{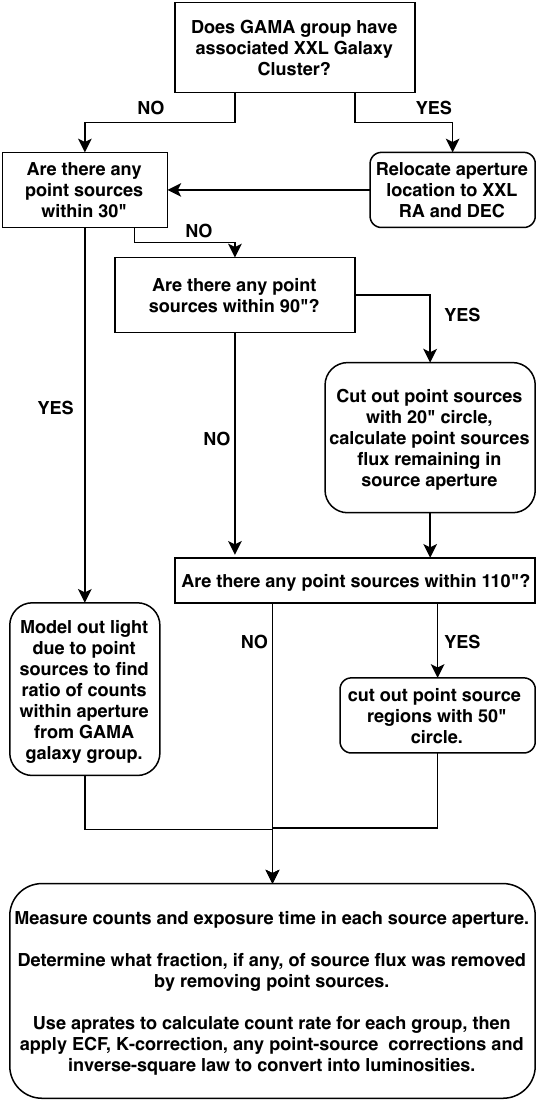}
    \caption[Flow chart of GAMA/XXL matches.]{A summary of the methodology used to conduct forced X-ray aperture photometry on the GAMA-XXL galaxy group sample. }
    \label{fig:matching_flowdiagram}
\end{figure}

Forced X-ray aperture photometry was used on the GAMA groups to allow the study of their X-ray properties in the regime of very low signal-to-noise X-ray data. 
While forced X-ray aperture photometry has been done on the XXL fields before \citep[e.g.][]{2018MNRAS.477.5517Willis}, the method used in this work differs slightly and is summarised graphically in Figure \ref{fig:matching_flowdiagram} with key points discussed here.

The first stage of the process was to match the GAMA and XXL group catalogues. 
The GAMA catalogue uses an iterative centre of light method to determine group positions, where the group's centre of light is calculated and the furthest galaxy removed until only two galaxies remain, from which the brightest is chosen as the group position.
This was used as an initial definition of the group centre when matching to the XXL catalogue.
The GAMA and XXL group catalogues were matched within 3000 km s\textsuperscript{-1}, with a GAMA group centre located within 2 Mpc and at least one member galaxy within 300 kpc of the X-ray location (in projection at the group redshift) \footnote{300 kpc corresponds to between 450" for the nearest groups to 50" for the most distant groups.}.
57 GAMA groups were found to have at least one matching XXL group, with 5 cases where a single GAMA group matched to two XXL groups.
For cases with two X-ray locations for a single GAMA group, we treated this as a single GAMA group, and the X-ray aperture photometry was conducted on both locations and then the luminosities were combined (in all cases the XXL $r_{500c}$ of both groups in such a pair overlap).
These cases will henceforth be referred to as "doubles".

While the overlapping region contains more GAMA than XXL detected groups out to the maximum GAMA redshift of the G02 sample ($z \leq 0.46$), there are 32 XXL groups which do not have GAMA counterparts in the GAMA-XXL sample used (when accounting for the 5 "doubles").
The lack of matches is due to our $\Nfof \geq 5$ cut.
All but one of the XXL groups had a GAMA counterpart when this threshold was lowered to $\Nfof \geq 3$.

For groups without an XXL match, the X-ray photometry was centred at the GAMA iterative centre.
If this location were significantly offset from the centre of the ICM distribution, this would lead to an underestimation of the X-ray luminosity.
Therefore, when there was a matching XXL group, the aperture was relocated to the XXL position, in order to better capture the emission from the ICM.
To test the impact of miscentring of the photometric apertures, the analysis was repeated using the GAMA location for all groups, regardless of any associated XXL detections.

A comparison of the subset of groups for which the aperture changed location (i.e. those where the aperture was shifted from the XXL to GAMA location) found that for the majority of groups, the X-ray luminosity did not change significantly when the aperture was shifted. 
We find that for 12 of 65 groups, changing the centre results in a change of luminosity at the 2$\sigma$ level. 
Based on this, we would expect $\approx20\%$ of the groups in our sample ($\approx47$ groups) to have luminosities that were underestimated due to miscentering.
The fraction could be higher for groups without an XXL detection as these tend to be lower mass systems with a less well defined centre.
To investigate the impact of using XXL locations, all of the analysis in this paper was repeated using apertures located at the GAMA centre for all groups, regardless of whether they have a matching XXL group, and we find it has no significant effect on our results.

The X-ray emission from the ICM in galaxy groups may be contaminated by point sources (mainly AGN) associated with the member galaxies, or projected onto the group.
Before proceeding with the X-ray photometry, these contaminating sources must be accounted for.

\subsection{Point source contamination}
\label{sec:AGN}

The XXL detection pipeline determines if an X-ray detection is a clear point source, and if so it is labelled "P1" (detailed in \PaperXXVII).
To determine which GAMA groups may be affected by point source contamination, all GAMA groups were matched with the 3XLSS catalogue (\PaperXXVII). 
The location and separation of any P1 point sources within 110" were noted.
As described in more detail below, a matching radius of 110" was used since the aperture radius used for the forced X-ray photometry of each group was 60" and the largest mask used to exclude any point sources had a radius of 50".

The 3XLSS catalogue also contains many unclassified sources (i.e. not C1, C2 or P1).
These unclassified sources may include X-ray detections of the ICM in the GAMA groups, which were too faint to be classified as C1 or C2, and hence should not be excluded from our photometric measurements.
However, some of these unclassified X-ray sources may be AGN, which should be excluded. 
This was tested by matching unclassified XXL sources with the ALLWISE AGN catalogue \citep{2018ApJS..234...23Allwiseagn}, and 109 unclassified XXL sources matched within 20" of an ALLWISE AGN.
These matched sources were then treated as X-ray detections of the AGN and excluded from any photometric measurements.
The 74 remaining unclassified XXL sources were not excluded from the photometric measurements, as they may be an XXL detection of the group emission.

In total, 117 GAMA groups were associated with a nearby point source (i.e. with an offset of $\leq110"$; see Table \ref{tab:NumberofGroups}).
For these groups, we accounted for the contaminating emission using several approaches depending on the separation of the point source from the group location.

\begin{figure}
    \includegraphics[width=\columnwidth]{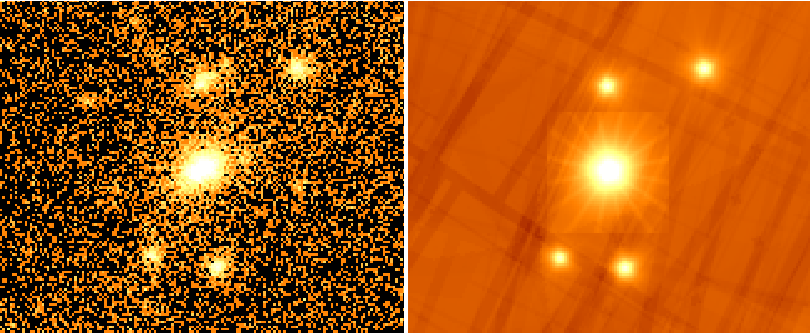}
    \caption[AGN within 30"]{Example of the Sherpa modelling of point sources within 30" of the group location. Left panel: \textit{XMM-Newton} image in the 0.5-2 keV energy band showing GAMA group 400098 with an XXL detected P1 source located at the GAMA group location. 4 further sources are located 86" to 136" away, and were included in the model. Right panel: The corresponding \texttt{sherpa} model, including instrumental effects, from which the cluster was calculated to contribute 15\% of the flux within a 60" aperture. Image size 300" x 300".}
    \label{fig:SherpaModel}
\end{figure}

For any group with a point source located within 30" of the group position, masking the source would exclude a large fraction of the ICM flux. 
Instead, the X-ray images of these groups were modelled using the CIAO\footnote{CIAO version 4.11 by \cite{Fruscione2006CIAO:System}} package \texttt{sherpa}.
A model of the ICM emission plus any point sources within a $300" \times 300"$ cutout centred on the group location was fitted to the X-ray image. 
The X-ray surface brightness distribution of the group was modelled using a 2d $\beta$-model \citep{1984ApJ...276...38J_Beta_model}, while point sources were modelled as delta functions.
The core radius, central position and amplitude of the $\beta$-model were free parameters in the fit, while the ellipticity of the model was fixed at zero and $\beta$ was fixed at a value of $2/3$. 
The positions and amplitudes of the delta functions were free to vary in the fit.
A flat background was assumed, with the normalisation a free parameter.
The model was multiplied by the appropriate exposure map and convolved with a model of the XMM PSF\footnote{PSF taken from Obs.ID 0037980101 observed on 11/01/2002 corresponding to MOS1 camera for an energy of 1keV. For detailed explanation about the \textit{XMM-Newton} PSF see \cite{XMM-PSF}}.
The X-ray flux measured for the group in our photometric analysis (with the point sources included) was then corrected for the proportion of the flux due to any point sources.
Figure \ref{fig:SherpaModel} shows an example of this modelling process for a GAMA group with a central X-ray point source and several other nearby sources.
In this particular case the modelling found that central point source was the dominant feature, and the ICM emission only accounted for 15\% of the flux within a 60" aperture.
However, in 16 of the 25 groups, the amplitude of the point sources were negligible in comparison to the group emission, indicating that the flux was dominated by the ICM.

For groups that did not have a point source within 30" of the centre, the sources were masked rather than being modelled in detail. 
Any point sources with a separation greater than 30", but less than 90", were removed by applying a circular mask of radius 20".
From our model of the PSF, the inner 20" contains 75\% of the flux from a point source.
The background subtracted flux within the 20" mask was used to estimate the point source flux that was outside the 20" mask but included in the 60" group aperture. 
This took into account the radial distribution of flux given by the PSF model.

Point sources located between 90" and 110" from the group centre were masked using a 50" circle which contains 90\% of the PSF and as such any residual contaminating within the group aperture was considered negligible.

In any cases where some region of the group aperture was excised to remove a point source, the ICM emission that was removed was corrected for using a 2d circular $\beta$-model.
For the purposes of computing the correction factor, the value of $\beta$ was fixed at $2/3$ and the core radius was fixed at $0.15 r_{500}$. 
Here, $r_{500}$ was calculated from the GAMA dynamical mass estimate of the group, and by scaling $r_{200}$ to $r_{500}$ (as described in \cite[\PaperIII]{XXLIII}).
The ratio of the flux under the $\beta$-model with and without a masked region at the location of the point source in question gave the correction factor to be applied to the measured group flux.
While the GAMA masses may have large errors (especially for groups with few members), this has a negligible effect on the correction factor, with a 50\% change in GAMA mass only having a 0.4\% change in the correction factor.

\subsection{Photometry details}
\label{sec:Photometry_discussion}

In this section we describe the method used to compute the posterior probability distributions for the luminosities of each group from their X-ray counts.
Due to limited data and the catalogue redshift range, a fixed aperture size of 60" about each group's location was used for the X-ray photometry.
Point source contamination was accounted for as discussed in Section \ref{sec:AGN}, and for 16 GAMA groups (i.e. 8 pairs), there was another GAMA group located within 120".
For these groups the overlapping region of the photometric aperture was excluded from the source counts for both groups, and the lost emission estimated using a $\beta$-model as discussed in Section \ref{sec:AGN}.
The background was calculated from a 2 deg$^2$ solid angle region around the group, with all point sources and galaxy groups masked.
This relatively large background region was chosen due to the nature of the data used, with large overlapping areas from the different CCD's and the large amount of the field which was masked.
The background was found to be relatively stable across the field ($2.5 \pm 0.1 \times 10^{-6}$ counts pixel$^{-1}$ s$^{-1}$).

The source count rate, $s$, was calculated using the CIAO package \texttt{aprates}\footnote{\url{https://cxc.cfa.harvard.edu/ciao/ahelp/aprates.html}}, as described by \cite{2014ApJ.Aprates}, where:
\begin{equation}
    \centering
    s = \frac{\frac{A_B T_B}{A_S T_S} n - m}{\frac{A_B T_B}{A_S T_S} T_S}, 
    \label{eq:aprates}
\end{equation}
with the parameters detailed in Table \ref{tab:Aprates}.
\texttt{aprates} returns a probability distribution function (PDF) for the count rate of each group, and is calculated using Poisson likelihoods for the source and background counts.

\begin{table}
    \centering
    \caption[Aprates input parameters]{Input parameters used by CIAO package \texttt{aprates} \citep{2014ApJ.Aprates} for equation \ref{eq:aprates}. 
    }
    \label{tab:Aprates}
    \begin{tabular}{ll}
        Parameter & Description \\
        \hline
        n & Source aperture counts \\
        m & Background aperture counts \\
        $A_s$ & Source aperture pixels \\
        $A_B$ & Background aperture pixels \\
        $T_S$ & Mean exposure time in source aperture \\
        $T_B$ & Mean exposure time in background aperture.
    \end{tabular}
\end{table}

\begin{figure}
    \includegraphics[width=\columnwidth]{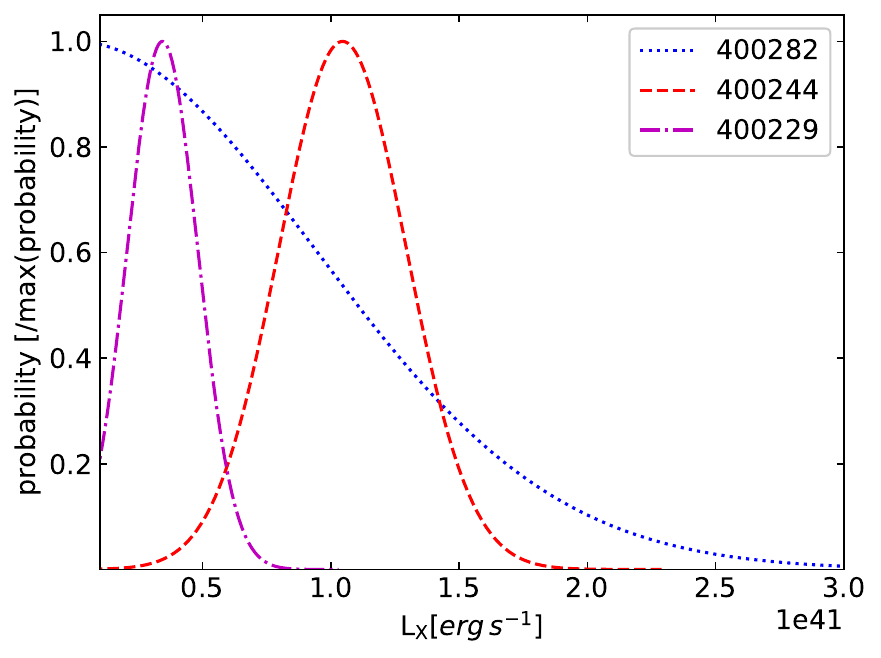}
    \caption{Example of the luminosity PDF for three groups. Group 400282 has a mode of zero, while groups 400229 and 400244 are approximately Gaussian and have well defined modes, although group 400229 is truncated. }
    \label{fig:TruncGauss}
\end{figure}

This photometry method has the advantage that each group's luminosity is represented by a PDF, even in the limit where there was no significant detection in the X-ray image.
For bright groups, the PDF will be approximately Gaussian, while for faint groups it may be truncated below or at the mode, examples of this are shown in Figure~\ref{fig:TruncGauss}.
Typically, analysis of a sample such as this, that contains faint and non-detections, would be based on luminosities that were a combination of measured values with errors and upper limits for non-detections. 
For our sample we found 58 groups with a mode luminosity of zero which would typically be treated as non-detections.
A significant benefit of our approach is that we make full and self-consistant use of the luminosity PDFs, thereby utilising all of the information available for formally undetected sources.

The count rate PDF for each group was then converted to an unabsorbed energy flux PDF, using an energy conversion factor (ECF), which takes into account the instrument response (in this case the equivalent of a single MOS exposure, consistent with the production of the mosaic), and an assumed source spectrum.
The ECF was calculated for each group using \texttt{sherpa} \citep{Freeman2001titleSherpa:Application/title}, assuming an APEC model \citep{Smith2001}, with Galactic absorption described by a phabs model \citep{Balucinska-Church1992PhotoelectricAbundances} with the absorbing column taken from the HI4PI survey \citep{BenBekhti2016HI4PI:GASS} at the position of each group.
The temperature of the APEC model for each group was estimated by applying the MT relation of \citet{Umetsu2020} to the GAMA group masses. 
However, this resulted in unrealistically low temperature estimates for some groups due to significant uncertainties in the low mass regime.
The ECF model used shows a steep increase for temperatures below 1 keV due to the large contribution from line emission at low temperatures, and the increased correction factors for the Galactic absorption and detector efficiency.
This introduces a strong dependency of the luminosity on the temperature assumed for cooler groups.
Previous work has found a threshold between the hot gas halos of individual galaxies and the group ICM at temperatures of $\sim1$ keV \citep[e.g.][]{Sanderson2003TheRelation}.
Therefore a minimum temperature of 1 keV was imposed.
 To test the effect of this assumed minimum temperature, the photometry was repeated with a minimum allowed temperature of 0.1 keV, and the impact on the results discussed in this paper was found to be negligible. 
 
The low signal to noise of the X-ray data for most of the groups prohibited measurements of the ICM abundance.
For this reason a fixed solar abundance of 0.3 Z\textsubscript{$\sun $} was assumed, using the abundance tables by \cite{Asplund2009TheSun}.
To investigate these assumptions, all of the analysis in this paper was repeated using the abundance tables of \cite{Anders1989AbundancesSolar} and also adjusting the abundance by $ \pm 0.1 $Z\textsubscript{$\sun $} from our standard value. 
None of these changes made a significant impact on the results discussed in this paper.

As the aperture size used was 60", a $\beta$-model was again used to extrapolate the luminosity out to $r_{500}$.
As in Section \ref{sec:AGN}, a constant value of $\beta$ = 2/3 was assumed, but there is evidence that lower-mass systems may have flatter surface brightness profiles.
For example, \citet{Sanderson2003TheRelation} suggests that for galaxy groups $\beta$ may be better represented as a power-law relation of the form $\beta = (0.44 \pm 0.06) T^{0.20 \pm 0.03}$.
Given the uncertainty in the temperature estimates (which are derived from GAMA dynamical masses), computing a temperature-dependent value of $\beta$ was not deemed useful. 
Instead, the photometry was repeated using $\beta=0.5$, corresponding to a $\approx 1$ keV system according to the \citet{Sanderson2003TheRelation} relation.
This typically increases the luminosity of a group by $\sim 15 \%$ when corrected to $r_{500}$.
As many groups have higher temperatures than 1 keV, this represents the upper end of the expected size of the systematic effect of flatter profiles. 
Again, the impact of this systematic on the results discussed in this paper was found to be negligible.
The uncertainty of the core radius used for aperture corrections, and hence the uncertainty on the final luminosity, also depends on the uncertainty on the GAMA dynamical mass estimates.
However, for a majority of groups this uncertainty is small relative to the width of the luminosity posterior.

\subsection{Luminosities}
\label{sec:Lums}

\begin{figure*}
    \includegraphics[width=0.8\textwidth]{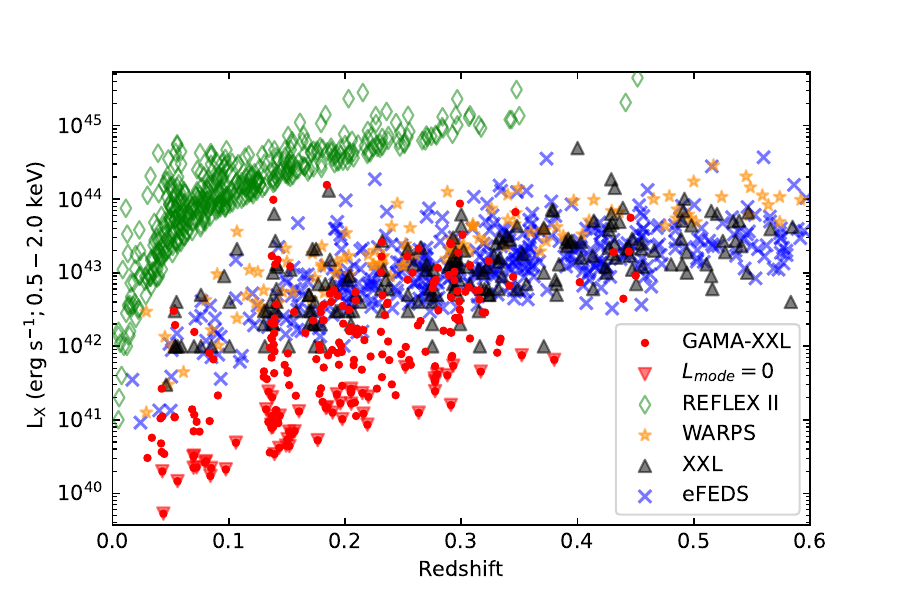} 
    \caption{Comparing the $0.5-2$ keV X-ray luminosities and redshifts of the GAMA-XXL sample with samples by REFLEX II \citep{REFLEXIV}, WARPS \citep{2008ApJS..176..374Horner_WARPS}, XXL (\PaperXX) and eFEDS \citep{Liu2022TheGroups}. The plotted redshift range is truncated at 0.6 to match the range of the GAMA sample but the X-ray selected samples extend to higher redshifts. The GAMA-XXL sample is represented by the median luminosity of each group, including those whose mode luminosity is zero which are highlighted by downwards triangles for clarity. REFLEX II luminosities were converted from 0-1-2.4 keV energy band using a correction factor of 0.6.}
    \label{fig:Luminosities}
\end{figure*}

To derive the final X-ray luminosities for the groups, the flux PDFs were k-corrected to the rest frame 0.5-2 keV band (using the same spectral model assumed for the ECF) and converted to luminosity PDFs.
In order to avoid unrealistic luminosity values, a lower limit of $10^{39}$ \ergps{} was applied to all luminosity PDFs.
This lower limit corresponds approximately to the predicted X-ray luminosity of a typical star-forming galaxy \citep{2012MNRAS.426.1870M}, which can be considered a conservative lower limit expected for the luminosity of the intra-group medium.

Figure \ref{fig:Luminosities} shows the luminosities of the GAMA groups plotted against redshift, alongside some other X-ray selected cluster samples.
For the purposes of this plot, the luminosity posteriors of the GAMA groups are represented by the median of the distributions. 
Note that these approximations are used for visualisation only; for our quantitative analyses we use the full posterior distributions.
The REFLEX II sample \citep{REFLEXIV} contains 802 bright X-ray selected clusters at $z\lesssim 0.5$ selected from the \textit{ROSAT} All-Sky Survey. 
The WARPS \citep{2008ApJS..176..374Horner_WARPS} sample comprises 124 clusters detected in pointed \textit{ROSAT} observations out to $z\sim 1$. 
The XXL sample (\PaperXX) lists the 365 brightest groups in the XXL survey, with known luminosities and redshifts for 235 of these groups out to $z\sim 1$.
Finally, the eFEDS \citep{Liu2022TheGroups} survey has detected 542 groups out to a redshift $z\sim 1.3$, detected using \textit{eROSITA}.
Note that the X-ray selected samples extend to $z\gtrsim1$, beyond the range included on this plot. 
The absence of GAMA groups at $z\gtrsim 0.4$ is a consequence of the optical selection. 
Figure \ref{fig:Luminosities} illustrates how the optical selection expands the parameter space to groups with X-ray luminosities below $10^{42}$ \ergps{} out to $z\approx 0.3$.

\begin{figure}
\centering
\includegraphics[width=0.9\columnwidth]{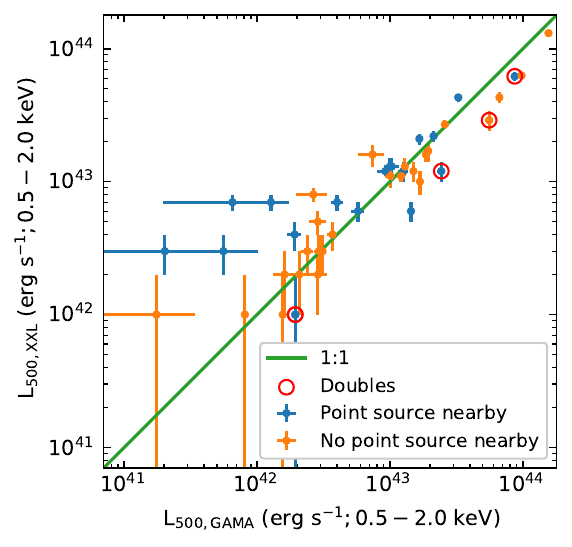}
\caption[Comparing mode of luminosity posterior with values for XXL detected groups]{Comparison between the median of each luminosity posterior with the values and errors given by \PaperXX{} for the GAMA groups which had XXL matches. The error bars on the GAMA luminosities show the 16\% to 84\% range of the luminosity posteriors. For the GAMA groups which had two XXL matches, the highest luminosity XXL match is used for comparison on this plot, marked as doubles. Only the 47 XXL groups which have luminosity constraints are included in this plot.}
\label{fig:modevsXXL}
\end{figure}

For the GAMA groups which matched to known XXL groups, we can compare our luminosities based on aperture photometry with those given by \PaperXX.
Given the high signal to noise of the XXL detected groups (relative to those detected by GAMA but below the XXL detection threshold), a more detailed analysis was used to calculate their luminosities \citep[\PaperII]{XXLII}.
This included the source temperature being measured directly, and a more bespoke background modelling (for groups in the XXL catalogue where the temperature was not well constrained, no luminosity value is given by XXL).
This level of analysis was not possible with the GAMA-XXL sample since the number of source counts too low for most of the groups.
To test the reliability of our simpler method, a comparison of the luminosities of groups in common to both the GAMA and XXL samples was conducted and is shown in Figure \ref{fig:modevsXXL}. 
On the whole, our X-ray luminosities are in agreement with those of \PaperXX.
Many of the cases in Figure \ref{fig:modevsXXL} where the GAMA luminosities are lower than those calculated by XXL might be explained by different treatments of point source contamination between the two analyses.

A number of groups were observed to have larger GAMA luminosities than those calculated by XXL.
Some cases are explained by doubles, where two XXL groups are classified as a single GAMA group according to the FoF algorithm.
These doubles are marked on Figure \ref{fig:modevsXXL}, where the brighter of the two XXL groups associated with the GAMA group is shown.
One of the doubles is not represented in Figure \ref{fig:modevsXXL} as neither XXL group had a X-ray luminosity constraint in the XXL catalogue.
As our method combines the luminosities from the two XXL locations, it would be expected that the GAMA luminosity is larger than that found by XXL. 

\PaperXLV{} used a similar procedure of forced aperture photometry (described by \citet{2018MNRAS.477.5517Willis}) to obtain $L_{300\text{kpc}}$\footnote{$L_{300\text{kpc}}$ is the luminosity within an aperture with a radius of 300 kpc.} values for GAMA groups in the XXL Field, which were found to agree well with the $L_{500}$ values used in this work, despite differences in methodology.
The comparisons with luminosities obtained by \PaperXX{} and \PaperXLV{} give validation to the forced X-ray aperture photometry methods used in this paper.

\subsection{Recovering Low Count Rates}
\label{sec:cr}

The GAMA-XXL sample used for this work is optically selected, and the X-ray flux for some groups is lower than the X-ray background level ($4.5 \pm 0.1 \times 10^{-3}$ counts s$^{-1}$ for a 60" aperture).
We tested the reliability of our method at recovering the count rate in this low signal regime using simple simulations of observed counts in a typical XXL observation with a realistic background and Poisson noise added.
We then used \texttt{aprates} to recover the posterior distribution for the count rate, and compared with the input value.

As can be seen in Figure \ref{fig:Overlap_Region}, the XXL North field contains a region of deeper exposure, from the XMM-Spitzer Extragalactic Representative Volume Survey \citep[XMM-SERVS,][]{Chen2018TheField} which targeted the original XMM-Large Scale Structure survey by \cite{Pacaud2007TheModelling}.
While the median exposure across the XXL-GAMA overlap region is $\sim$50 ks\footnote{Exposure times quoted at equivalent depth of a single MOS exposure.}, the median exposure time in the XMM-SERVS region is $\sim$150 ks, and the median exposure time excluding the XMM-SERVS region is $\sim$ 40 ks.
In order to obtain a conservative estimate of the count rate which \texttt{aprates} can recover, the test was conducted using an exposure time of 40 ks.

\begin{figure}
\centering
\includegraphics[width=\columnwidth]{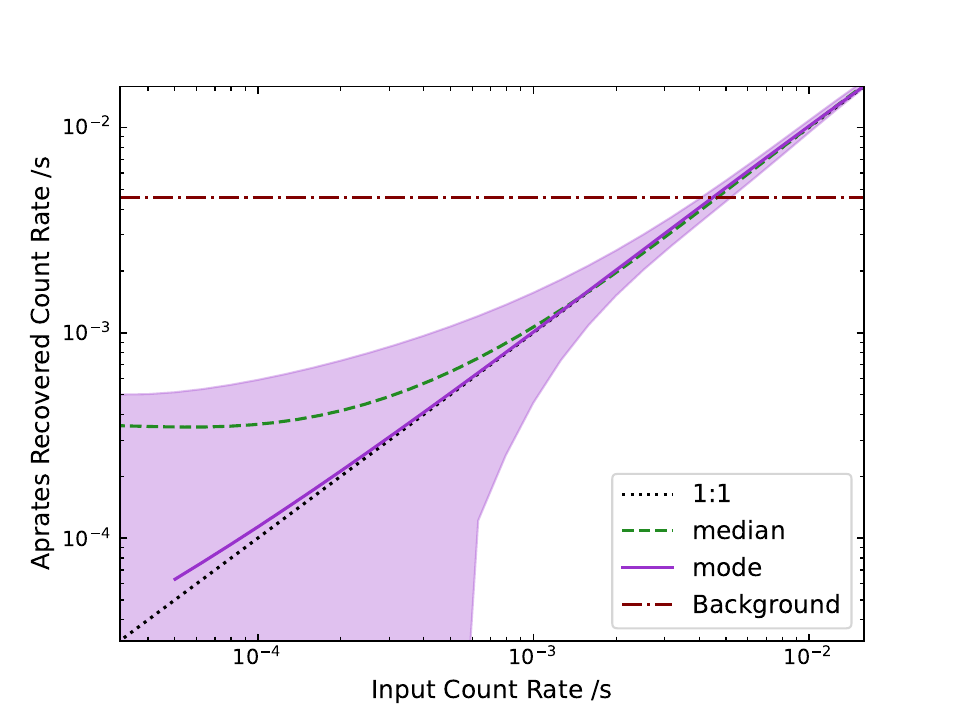}
\caption[Aprates Cr in vs cr out]{The count rate recovered by \texttt{aprates} over a range of input count rates for a 40ks exposure to represent the shallower XXL observations. The test was done for a 60" source aperture with a background count rate of $4.5 \times 10^{-3}$ counts s$^{-1}$ included before taking a Poisson realisation for the source and background apertures. For each input count rate, 100 Poisson realisations were taken and the mode from each output was used to build up the \texttt{aprates} recovered count rate. The median and standard deviation of the 100 recovered mode count rates for each input count rate is indicated by the purple line and shaded region. Each output count rate posterior was also sampled 50 times to represent the median output count rate (green line).}
\label{fig:crin_crout}
\end{figure}

Figure \ref{fig:crin_crout} summarises the results of 100 Poisson realisations for each input count rate.
For the purposes of this comparison we summarise the posteriors by using their mode and median.
As can be seen in Figure \ref{fig:crin_crout}, at low count rates the mode and median diverge as the posteriors become more asymmetric, resulting in the median becoming a biased estimate of the count rate below $\approx 1 \times 10^{-3}$ counts s$^{-1}$.
The good agreement between the mode and input count rate demonstrate that the posteriors are unbiased albeit increasingly uninformative for count rates down to $1 \times 10^{-4}$ counts s$^{-1}$. 
Below this, the majority of simulations return a mode-zero posterior, as demonstrated by the termination of the mode line in Figure \ref{fig:crin_crout}.

While the analysis done in this paper utilises the full posteriors, for cases where the luminosity of a group needed representing by a single value (e.g. for visualisations such as Figure \ref{fig:Luminosities}) we use the median of the posterior.
While this is a biased representation of the luminosity for the lowest count rate groups, the mode would not have been particularly useful due to number of groups with mode-zero posteriors.

\section{The GAMA group X-ray Luminosity Function}
\label{sec:XLF}

The XLF is the comoving number density (\textit{n}) of objects per luminosity interval, which decreases following a power-law for low luminosity objects and exponentially for high luminosity objects.
This is represented by the Schechter function \citep{1976ApJ...203..297Schechter}:
\begin{equation}
    n(L_x)dL_x = n_0 \left(\frac{L_x}{L^{*}_{x}}\right)^{-\alpha}\exp{\left(\frac{-L_x}{L^{*}_{x}}\right)}\frac{dL_x}{L^{*}_{x}}
    \label{eq:Schechter}
\end{equation}
where $n_0$ normalises the XLF, and $\alpha$ determines the power-law slope when the X-ray luminosity ($L_x$) is less than $L^{*}_{x}$.

The vast majority of the GAMA groups are fainter than the characteristic luminosity $L_X^*$ typically found ($\sim 10^{44}$\ergps)and so $L_X^*$ cannot be constrained by our data and is fixed as $2.59\times 10^{44}$ \ergps \citep{2013MNRAS.435.3231Koens_WARPS}.
The power of our sample is in constraining the faint end of the XLF at much lower luminosities than previous studies, and the precise value adopted for $L^*$ does not have an impact on our results.
In the low-luminosity limit covered by our data, the Schechter function is approximately a power-law, with a pivot at $L_X^*$.
Since our data are all below the pivot luminosity, this introduces a strong degeneracy between $n_0$ and $\alpha$.
To remove this we rewrite equation \ref{eq:Schechter} to be normalised at $10^{42}$ \ergps, such that the pivot luminosity is closer to the median luminosity of the sample.
Therefore, the modified Schechter function follows the form:
\begin{equation}
    n(L_x)dL_x = n_{42} \left(\frac{L_x}{L_{42}}\right)^{-\alpha}\exp{\left(\frac{-L_x}{L^{*}_{x}}\right)}\frac{dL_x}{L^{*}_{x}},
    \label{eq:Schechter2}
\end{equation}
where
\begin{equation}
    n_{42} = n_0\left(\frac{L_{42}}{L^{*}_{x}}\right)^{-\alpha} .
    \label{eq:n42}
\end{equation}

In a conventional representation of the XLF, luminosity bins are constructed and the number of clusters in each bin are computed. 
In our analysis, the luminosity posteriors may extend across multiple luminosity bins.
For this reason, the number of objects in a given luminosity bin was calculated by summing the fraction of the posterior of each GAMA group within that bin.
The contribution of a group to a given bin is given by the integral of its posterior across that bin.
The luminosity bins were set to be of equal width in log space (0.2 dex) over the range $ 38 \leq \log_{10}(\text{L}_{500} / \text{\ergps}) \leq 44.2 $

The number density of groups in a luminosity bin was calculated using an adapted version of the $1/V_{max}$ method \citep{Schmidt1968SpaceSources, Avni1980OnQuasars}.
If $r_i$ is the integral of the luminosity posterior of group $i$ over luminosity bin $j$ of width $\Delta L_j$ (i.e. the probability of group $i$ belonging to bin $j$), then the number density of groups in that bin is given by
\begin{equation}
    n_j = \frac{1}{\Delta L_j} \sum_i^N \frac{r_{i}}{V_{\text{max},i}}.
    \label{eq:n}
\end{equation}
Here $V_{max,i}$ is the maximum comoving volume at which the $i$th group can be observed.
The maximum comoving volume was defined as the volume by the redshift, $z_{\text{max},i}$, at which the 5th brightest member in the group would have an apparent r-band magnitude of 19.8, the GAMA magnitude limit used in constructing the group sample.
When determining $z_{\text{max}}$, a k-correction was applied to the \textit{r}-band magnitude of each galaxy using \textit{KCORRECT} v5.1.3 \citep{Blanton2007Near-Infrared} and the smooth spectral energy distribution coefficients estimated by \citet{Loveday_2010_GAMA_GalaxyLF} \footnote{\url{https://www.gama-survey.org/dr4/data/cat/kCorrections/}}.

\begin{figure*}
    \includegraphics[width=1.6\columnwidth]{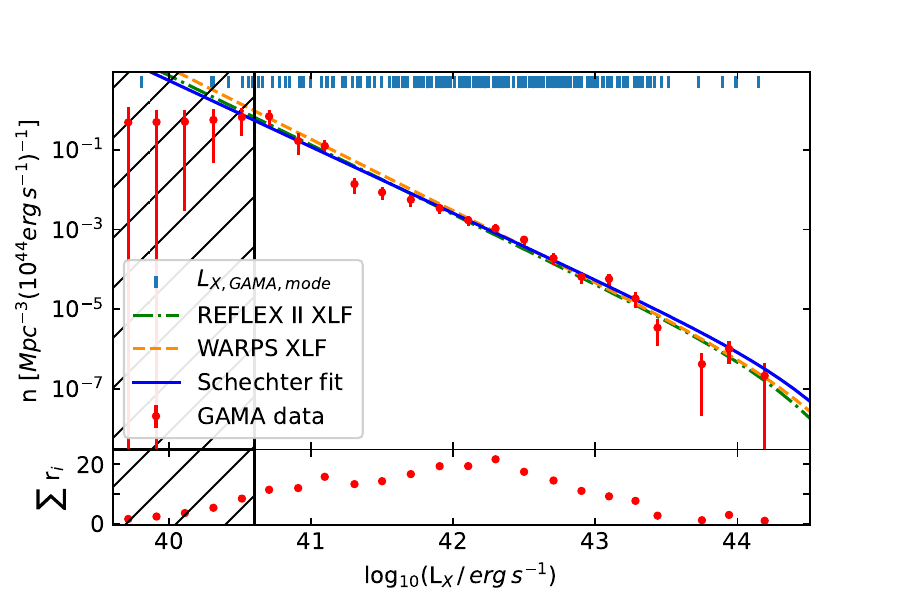} 
    \caption{The XLF of the the optically selected GAMA sample along with the best fitting Schechter function, $n_{42} = (6.63\pm0.61) \times10^{-3} \text{Mpc}^{-3}$ and $\alpha = 1.66\pm0.05$, with $L^*$ fixed at $2.59\times 10^{44}$ \ergps. 
    The fit to the GAMA data was limited to luminosities $\gtrsim 10^{40.6}$ \ergps{} due to bins below this being dominated by mode zero groups.
    The excluded bins are shown by the hatched region, the mode luminosities are represented by blue dashes along the top of the plot (excluding the 58 groups with mode-zero luminosity posteriors). 
    Schechter functions found by REFLEX II \citep{REFLEXIV} and WARPS \citep{2013MNRAS.435.3231Koens_WARPS} are included for comparison and show good agreement, even when extrapolated below the $\gtrsim 10^{42}$ \ergps{} luminosity ranges used for those works.
    The lower panel shows the effective number of groups in each bin (i.e. the sum of the integral of each groups luminosity posterior located within a given bin.}
    \label{fig:XLF}
\end{figure*}

The resulting XLF is shown in Figure \ref{fig:XLF}.
The luminosity posteriors within each bin were summed to produce a total probability density, the mean of which was then taken to represent the luminosity of that bin.

The uncertainty, $\sigma_j$, on the number density of groups in a given bin results from the uncertainty on the luminosity of each group that contributes towards that bin, and the Poisson uncertainty on the number of groups expected in the bin for a given number density.
The uncertainty on $n_j$ due to the luminosity uncertainty was estimated using a Monte-Carlo approach. 
A random realisation of the XLF was generated by sampling a luminosity from the posterior of each group and assigning each group to the corresponding luminosity bin. 
This process was repeated 1,000 times and $\sigma_{n,j}$ was calculated as half of the difference between the 16th and 84th percentiles of the number density in each luminosity bin. 
The uncertainty due to the counting statistics on each bin was calculated as the square root of the sum of $r_i$ in that bin. 
The error on the number density in bin $j$ is then given by
\begin{equation}
    \sigma_j = \sqrt{\sigma_{n_j}^2 +\left(\frac{n_j}{\sqrt{N_j}}\right)^2}.
    \label{eq:error}
\end{equation}
This method accounts for the fact that the number densities in different bins are not independent due to the contribution of the luminosity posteriors across multiple bins.
The contribution to the uncertainty from each term was roughly equivalent, with $\sigma_{n,j}$ contributing more in the lower luminosity bins and the uncertainty due to counting statistics dominating in the highest luminosity bins.

A Schechter function was fit to the XLF data shown in Figure \ref{fig:XLF} using a Bayesian approach.
The likelihood of the data given the model was calculated assuming a log-normal probability of observing $n_j$, given the number density predicted for that bin by the Schechter function, and the uncertainty $\sigma_j$.
The prior on $n_{42}$ was uniform in log space between $10^{-8}$ Mpc\textsuperscript{-3} and $10^{-5}$ Mpc\textsuperscript{-3}, and the prior on $\alpha$ a Student’s t distribution with 1 degree of freedom, which is mathematically equivalent to a uniform prior on the angle of the line, as a uniform prior on the gradient would be weighted towards larger angles.

Eddington bias originates from the scatter in the property of a population whose number density changes greatly between adjacent bins of said property.
For a volume complete sample, there will be more objects scattered into a given bin from bins with higher number densities, than objects scattered out of the bin.
The GAMA-XXL sample used in this work is not volume complete, but apparent magnitude complete, which negates the impact of Eddington bias.
This is because fainter groups, while having a higher number density than brighter groups, have a smaller detection volume.
From the lower panel in Figure \ref{fig:XLF} we see that the effective number of groups in each bin does not deviate greatly between bins.
Therefore Eddington bias shouldn't be a factor in the results, as scatter in luminosity would move a similar proportion of groups in both directions.
Although the number of groups in each bin remains fairly level, the number density contribution of groups follows a power-law because of the V$_{max}$ correction applied.
Therefore, the scatter in luminosity would preferentially redistribute the larger contribution in number density from low luminosity bins to high luminosity bins.
Our resampling method (explained above) means that the impact of the scattering of number density contributions is included in our uncertainties.

The mode of the luminosity posteriors of each group are shown in Figure \ref{fig:XLF} as ticks along the top axis, with groups with a mode of zero not shown due to the logarithmic scale. 
It is clear that the lowest luminosity bins are dominated by "mode-zero" groups.
For bins below $10^{40.6}$ \ergps, over half of the density comes from mode-zero groups and so these bins are excluded from any further analysis as the number density in each bin cannot be well constrained. 
The observed flattening in these low luminosity bins is discussed in Section \ref{sec:N>5}, and shown to not impact bins above the $10^{40.6}$ \ergps{} threshold applied when fitting.
We note that the mode-zero groups still contribute to the XLF as their luminosity posteriors extend into the higher bins, and so they are included as $r_i$ in equation \ref{eq:n}.
The summed integrals of each group's luminosity posterior within the fitted area adds up to 211, and so the equivalent of 24 groups are excluded from the fitting due to the cut at $10^{40.6}$ \ergps.
While the main results in this paper will be discussed in the context of this luminosity cut, in section \ref{sec:Lcuts} we will investigate the impact of the different luminosity cuts which could be applied.

\begin{figure}
    \includegraphics[width=\columnwidth]{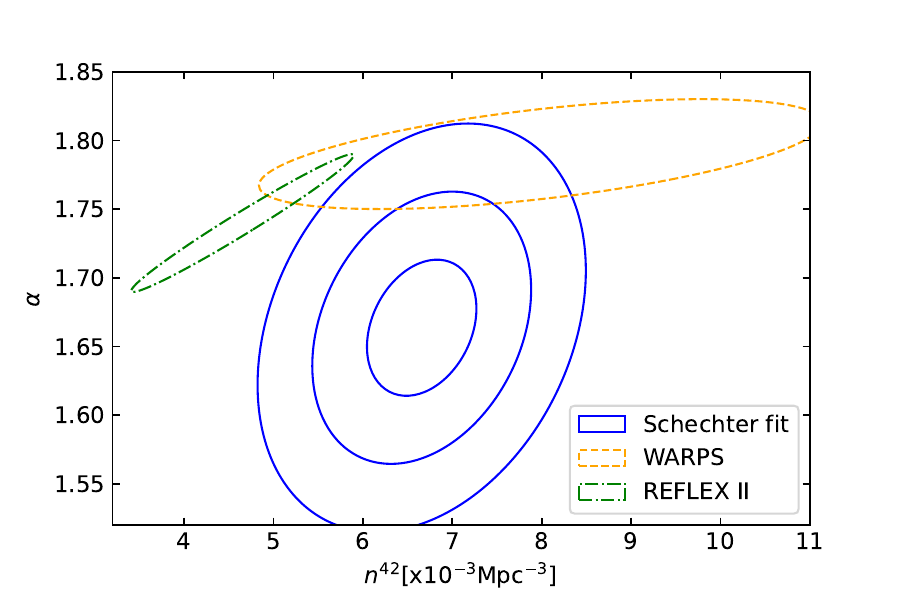} 
     \caption{The posterior distribution of $n_{42}$ and $\alpha$ for the data shown in Figure \ref{fig:XLF}. Contours indicate $1\sigma$, $2\sigma$, and $3\sigma$ confidence intervals. $1\sigma$ contours from REFLEX II \citep{REFLEXIV} and WARPS \citep{2013MNRAS.435.3231Koens_WARPS} included for comparison, having been converted to the form given by equation \ref{eq:Schechter2}.  }
    \label{fig:XLF_posterior}
\end{figure}

The posterior probability of the XLF parameters $n_{42}$ and $\alpha$ was sampled using \texttt{emcee} \citep{Foreman-Mackey2013} and is shown in Figure \ref{fig:XLF_posterior}.
The resulting best fitting values for the Schechter function are $n_{42} = (6.63\pm0.61) \times10^{-3} \text{Mpc}^{-3}$ and $\alpha = 1.66\pm0.05$, and are illustrated in Figure \ref{fig:XLF}.
While our model was parametrised in terms of $n_{42}$ to avoid the degeneracy between $n_{0}$ and $\alpha$, for convenience we calculate a value of $n_0 = (6.40 \pm 0.61) \times10^{-7}$ Mpc$^{-1}$, but note that the uncertainties on $n_0$ are correlated with those on $\alpha$.

Figure \ref{fig:XLF} also shows the best-fitting Schechter functions found by REFLEX II \citep{REFLEXIV} and WARPS \citep{2013MNRAS.435.3231Koens_WARPS} for comparison.
For figure \ref{fig:XLF_posterior}, the best-fitting parameters were converted from $n_0$ to $n_{42}$ using equation \ref{eq:n42}.
This conversion is dependant on the slope of the XLF in question and so introduces a degeneracy between $n_{42}$ and $\alpha$ which is apparent in Figure \ref{fig:XLF_posterior}.
This conversion was done by sampling values of $n_0$ and $\alpha$ at $L^\star_X$, assuming they are independent and using the reported errors.
Each $n_0$, $\alpha$ pair was then used to convert $n_0$ to $n_{42}$.
The XLF by \cite{REFLEXIV} was measured in the 0.1 - 2.4 keV energy band and so we applied a multiplicative factor of 0.6 to the normalisation to correct to the 0.5 - 2.0 keV energy band.
The GAMA XLF is in good agreement with the X-ray selected samples in the luminosity range probed by those samples ($\gtrsim 10^{42}$ \ergps).

\begin{figure}
    \includegraphics[width=\columnwidth]{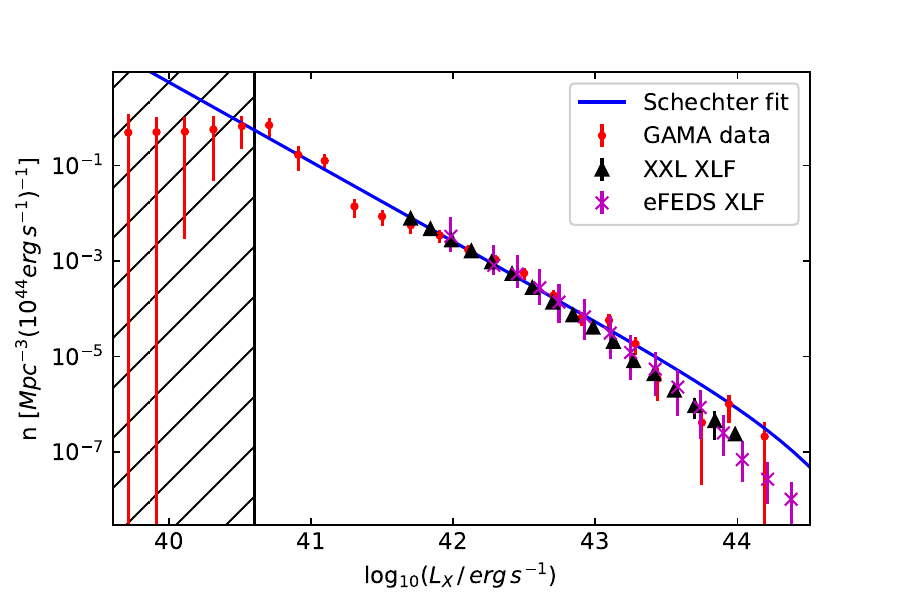} 
    \caption{Comparison of the GAMA-XXL sample XLF and best-fitting Schechter function (as in Figure \ref{fig:XLF}), with the XLFs found by the XXL and eFEDS surveys \citep[\PaperXX, ][]{Liu2022TheGroups}.
    }
    \label{fig:XLF_litcomp}
\end{figure}

In Figure \ref{fig:XLF_litcomp} we further compare the GAMA XLF with the binned data from some recent X-ray surveys: XXL (\PaperXX) and eFEDS \citep{Liu2022TheGroups}.
The XXL XLF is consistent with ours in the high luminosity regime ($\gtrsim10^{42}$ \ergps), which is expected given that they are derived from the same X-ray data (different photometric methods notwithstanding).
Similarly good agreement is seen with the eFEDS data above $10^{42}$ \ergps. 
We note that the number density of groups in the highest luminosity bins tend to fall below that expected from the best fitting Schechter functions derived by this work or from REFLEX II and WARPS.
This could have interesting implications on the choice of $L^\star_X$, however, this is beyond the scope of this work due to the lack of high luminosity groups. 
The key difference is the extension of the GAMA XLF to lower luminosities by over an order of magnitude, enabled by the optical selection.

\section{The Luminosity-Mass Relation}
\label{sec:LM}

While the XLF is a useful description of an observable property of galaxy groups, the halo mass function (HMF) is the fundamental description of the cluster population. 
Much effort continues to be put into constraining the HMF observationally in order to constrain cosmological parameters. 
If the relationship between X-ray luminosity and halo mass (the LM relation) were known, then our observed XLF could be used to measure the HMF. 
In fact, the LM relation of low-mass groups is poorly constrained and the intrinsic scatter ($\sigma_L$) large, therefore it is more informative to assume a fixed cosmology and use our data to constrain the LM relation of our sample. 
In effect, we will infer the form of the LM relation required to map the observed XLF onto a known HMF.
For this work we will assume the HMF fitting functions derived by \cite{2008ApJ...688..709Tinker_Massfunc}.
The methodology used is similar to that by \cite{Mantz2010TheConstraints}, who found that the results are indifferent to the choice of mass function.
Using GAMA data, \cite{Driver2022AnData} reconstructed the HMF empirically down to $M_h = 10^{12.7} M_\odot$ showing broad agreement with the expectation from \cite{2008ApJ...688..709Tinker_Massfunc}.

We adopt a power-law description of the LM relation:
\begin{equation}
    \label{eq:LM}
    L_x = A \, E(z)^C \left(\frac{M}{M_0}\right)^B
\end{equation}
where $M_0 = 5\times10^{13} \, \text{M}_{\odot}$. 
$E(z)=\sqrt{\Omega_\text{M}(1+z)^3 + \Omega_\Lambda}$ describes the redshift-dependence of the Hubble parameter, and we fix the exponent $C=2$, corresponding to self-similar evolution for our luminosity energy band \citep{Ettori2015TheProfile}. 
We note that due to the low-redshift nature of the sample, the impact of our assumed evolution is negligible.
Since each luminosity bin in the XLF has contributions from a large number of groups at different redshifts, we used the median redshift ($0.196$) for our calculations.

To find the best-fitting parameters for equation \ref{eq:LM}, we use a forward modelling technique to predict the XLF from the HMF for a given LM relation, and then compare the resulting prediction with the observed XLF.
For a given pair of values of $A$ and $B$, the mass corresponding to each luminosity bin in Figure \ref{fig:XLF} was calculated.
For each mass, the number density per unit mass ($dn/dlnM$) was computed using the python library Colossus \citep{Diemer2018COLOSSUS:Halos}.
This was then converted to a number density per unit luminosity ($dn/dlnL$) using equation \ref{eq:LM}.
In order to incorporate the scatter in the LM relation, the value for $dn/dlnL$ in each luminosity bin is the sum of the expected contributions from all overlapping bins given the value of $\sigma_L$.
The likelihood of the observed XLF given this model was then computed assuming lognormal errors on the observed number densities (as in Section \ref{sec:XLF}). 
As before, the fit and main results in this paper were limited to points with luminosities $\gtrsim 10^{40.6}$ \ergps.
In section \ref{sec:Lcuts} we will investigate the impact of the different luminosity cuts which could be applied.

The approach taken to incorporate the scatter leads to a strong degeneracy with normalisation due to the power-law shape of the HMF.
Larger values of $\sigma_L$ result in a greater contribution to the value of $dn/dlnL$ in a given luminosity bin from lower luminosity bins, leading to a decrease in the normalisation required for the LM relation.
The ability to constrain $\sigma_L$ is limited by the bin width used for fitting the XLF and the degeneracy with the normalisation.
As a result, we adopt a Gaussian prior $\mathcal{N}(0.2,0.1)$ for $\sigma_L$
which is a conservative estimate, with other studies finding values up to 0.5 \citep[e.g.,][]{2007ApJ...668..772Maughan_LM, Andreon2016TheSample}, and so the normalisation value calculated in this paper can be considered an upper limit.

\begin{figure}
    \centering
    \includegraphics[width=\columnwidth]{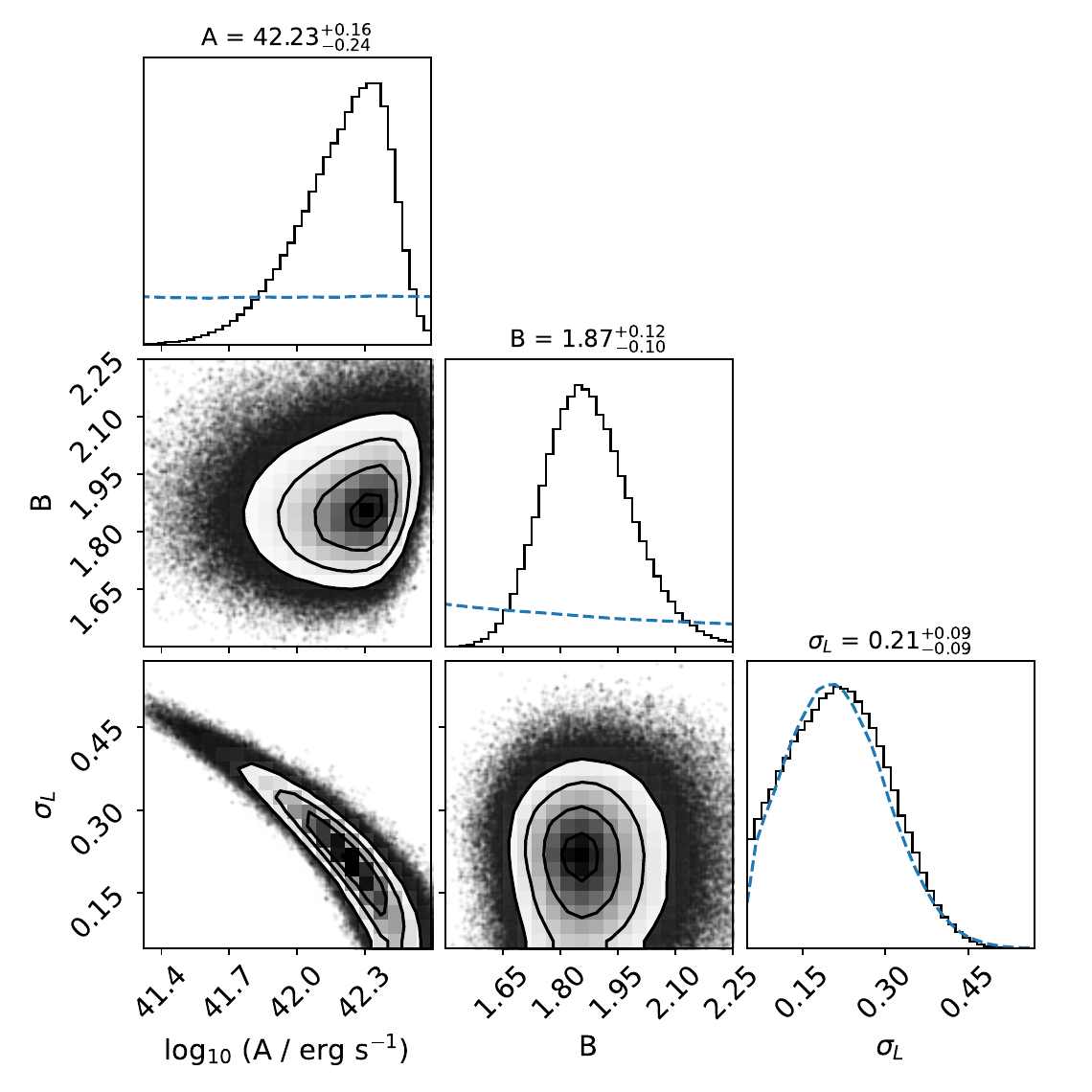}
    \caption{The posterior distribution for the normalisation, slope and scatter of equation \ref{eq:LM} required to reproduce the observed XLF of the GAMA groups given a theoretical HMF. 
    The blue dashed line represents the priors used, illustrating that the posterior distribution of $\sigma_L$ closely resembles the Gaussian prior $\mathcal{N}(0.2,0.1)$. 
    The quoted values represent the median with the upper and lower errors calculated from the 16th and 84th percentiles. 
    The 2d contours for the normalisation and slope are shown in Figure \ref{fig:LM_AB} with a comparison to other studies.
    The bottom left panel demonstrates the strong degeneracy between the normalisation and scatter of the LM relation. 
    Since the prior used for $\sigma_L$ in this work is a conservative estimate, the normalisation value can be considered an upper limit.}    
    \label{fig:LMcorner}
\end{figure}

We adopted a prior on $A$ that was uniform in log space between $0.01\times 10^{43}$ \ergps{} and $3\times 10^{43}$ \ergps, and the prior on $B$ was a Student's t distribution with 1 degree of freedom, which is mathematically equivalent to a uniform prior on the angle.
The resulting posterior for the LM relation was sampled using \textit{emcee} and shown in Figure \ref{fig:LMcorner} along with the priors.
The resulting best fitting values are $A = (0.17 \pm 0.07) \times 10^{43}$\ergps{} and $B = 1.87 \pm 0.12$, with a comparison to other studies discussed in Section \ref{sec:discuss_LM}.

The form of the LM relation derived in this paper is limited by the accuracy and bin size of the XLF.
A more rigorous method would be to use a complete Bayesian approach, and determine the Poisson likelihood of the observed cluster counts given expectations from a mass function when scaling relations and selection effects are applied.
However, modelling the appropriate selection function is beyond the scope of this paper due to poor constraints on group properties in the low mass regime.

The method used depends on the chosen cosmology, as the normalisation of the HMF is sensitive to the choice to $\sigma_8$.
For this work the WMAP9 cosmology by \citet{Hinshaw13} was assumed, however, to test this choice the entire analysis (luminosity calculation, XLF fitting and LM relation fitting) was repeated using the Planck18 cosmology by \citet{Planck2018a} with $H_0 = 67.66$, $\Omega_M$ = 0.3111, $\Omega_R$ = 0, $\Omega_\Lambda$ = 0.6888, and $\sigma_8 = 0.81$.
Using Planck18 cosmology the best fit parameters for the LM relation are $A = (0.12 \pm 0.04) \times 10^{43}$ \ergps and $B = 1.76 \pm 0.1$, which are within $1 \sigma$ of the values found by WMAP9 cosmology, and the posteriors are comparable to that shown in Figure \ref{fig:LMcorner} for the GAMA-XXL sample.

\section{Discussion}
\label{sec:Discussion}

We've constructed the XLF and inferred the form of the LM relation for an optically selected sample of groups, which extends over an order of magnitude fainter than is typically reached in X-ray selected samples.
This gives us an opportunity to probe the impact of non-gravitational processes on the ICM properties in these low mass systems.
In the following, we discuss the possible systematics impacting the GAMA-XXL sample and our ability to constrain the low-luminosity groups, before comparing our results with other studies and consider their astrophysical interpretation.

\subsection{GAMA completeness}
\label{sec:N>5}

The results presented in this work depend on the reliability in our modelling of the GAMA selection function, which is our imposed $\Nfof \geq 5$ richness limit. 
A possible concern is that for a given luminosity bin in the XLF, there could be GAMA groups whose X-ray luminosity is located within that bin, but that were excluded due to having $\Nfof<5$.
This would then lead to an underestimate of the number density of groups in that luminosity bin.
However, a GAMA group with $\Nfof=4$ can be thought of as a GAMA group with $\Nfof=5$ that is too far away for the fifth brightest galaxy to be brighter than the GAMA magnitude limit.
This is precisely the effect that the $V_{max}$ method is intended to take into account.

\begin{figure}
    \includegraphics[width=\columnwidth]{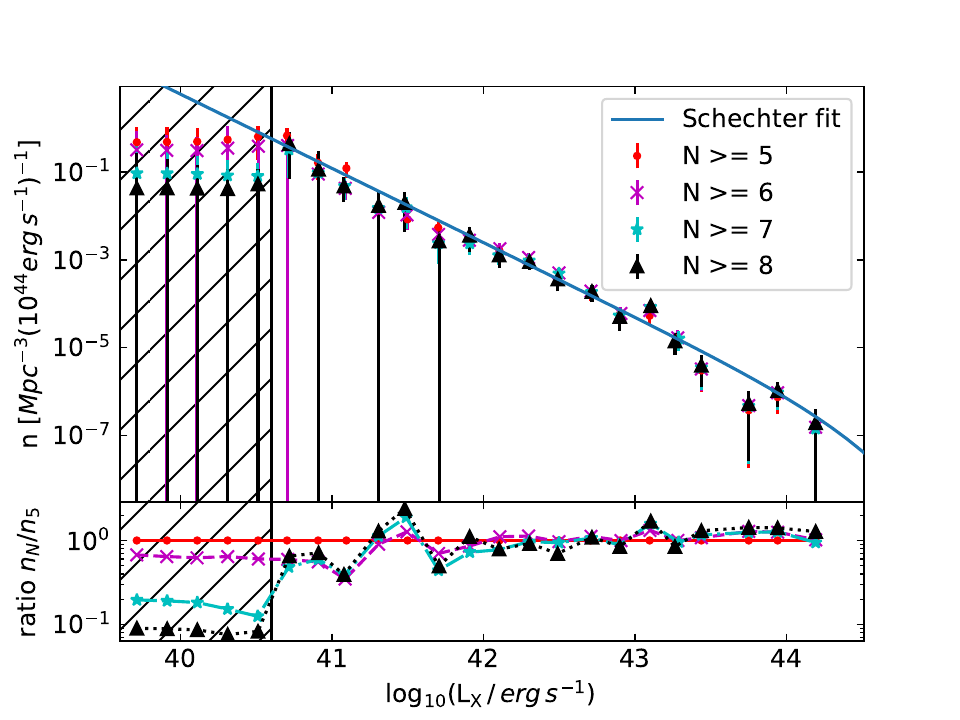} 
    \caption{Comparing the number densities for different GAMA richness limits shown in Table \ref{tab:N5678}. The $\gtrsim 10^{40.6}$ \ergps{} cutoff and Schechter fit are as described in Figure \ref{fig:XLF}, while the bins width is increased to ensure a reasonable number of groups in each bin for the higher richness limits.}
    \label{fig:XLF_N5678}
\end{figure}

To test whether the $V_{max}$ method correctly models our selection function, we repeated the XLF analysis with increasing thresholds applied to $\Nfof$.
If the $V_{max}$ correction were not effective, then raising the selection threshold should have the effect of lowering the number density of groups.
The $\Nfof$ limits used and the relevant sample sizes are detailed in Table \ref{tab:N5678}.
The XLF was re-calculated for each subset, with the volume for each group adjusted to reflect the selection threshold for that subset.
Increasing the $\Nfof$ threshold reduces the sample size, and also reduces the $V_{max}$ for each group, recovering an approximately constant number density in each luminosity bin, as intended.
This is shown in Figure \ref{fig:XLF_N5678}, which demonstrates that there is no systematic effect on the number densities caused by the changing selection function above the luminosity threshold used for fitting the XLF ($10^{40.6}$ \ergps).

The approach taken here follows that done by \citet{Driver2022AnData} who found that the onset of incompleteness for the GAMA sample is at a mass of $10^{12.8} M_{\odot}$.
Using our best fit LM relation, this corresponds to a luminosity of $10^{40.6}$ \ergps, and for bins below this threshold there is an apparent dependence to the number density on the choice of the $\Nfof$ cut.
This implies that the dependence may be in part due to the onset of incompleteness in the GAMA sample for groups below this threshold.
This dependence is also in part contributed to by the uninformative nature at the low luminosity end of the posterior of mode-zero groups.
This is due to the normalisation of the XLF in these low luminosity bins being dependant on the lower limit applied to the luminosity posteriors.

It would be expected that the lower limit of the luminosity of the sample would scale with the lower limit on $\Nfof$, therefore, the normalisation decreases with the choice of $\Nfof$ cut as the lower luminosity limit remains constant.
This also helps confirm that for bins above $10^{40.6}$ \ergps, the number density is independent to the lower luminosity limit applied to the luminosity posteriors.
It could also be expected that we would observe some some Eddington bias in the $\Nfof$ cut applied to the GAMA group sample.
This would arise as there are more low mass groups, and so more groups for which the scatter will raise the $\Nfof$ value above the selection threshold than decrease below the threshold.
If this Eddington bias had an impact on the results, then increasing the selection threshold would result in a drop in the number density of every higher luminosity bins.
Figure \ref{fig:XLF_N5678} shows that a stricter selection threshold has no systematic effect on the number densities for bins above $10^{40.6}$ \ergps.
Thus, we conclude that the $V_{max}$ method adequately models our GAMA selection above the luminosity threshold used for fitting the XLF.

\begin{table}
    \centering
    \caption{The sample size for each selection limit used when testing the modelling of the GAMA selection function.}
    \label{tab:N5678}
    \begin{tabular}{ll}
         $\Nfof$ limit & Sample size \\
         \hline
         $\geq5$ & 235 \\
         $\geq6$ & 163 \\
         $\geq7$ & 123 \\
         $\geq8$ & 95 \\
    \end{tabular}
\end{table}

\subsection{Low Luminosity Constraints}
\label{sec:Lcuts}

Due to the GAMA-XXL sample used for this work being optically selected, we expect the X-ray flux for some groups to be below the X-ray background level.
This results in poorly constrained X-ray luminosity posteriors for some groups (e.g. Figure \ref{fig:TruncGauss}).
To account for "mode-zero" groups, the results discussed in this paper only include luminosity contributions to bins above $10^{40.6}$ \ergps.
This is due to bins below this luminosity limit being dominated by "mode-zero" groups, as discussed in Section \ref{sec:XLF}.
Here we will investigate the impact of a more conservative luminosity cut on the results.

Section \ref{sec:cr} shows that \texttt{aprates} can recover an informative posterior for the count rate of a source down to below 1 $\times 10^{-3}$ counts s$^{-1}$.
This is a conservative estimate calculated for the region of low exposure.
For a group located at the sample's median redshift (0.196) this corresponds to a luminosity of $10^{41.75}$ \ergps.
Applying this luminosity cut to the XLF fitting procedure described in Section \ref{sec:XLF} leads to a slight increase in the normalisation and slope ($n_{42} = (7.6\pm1.2) \times10^{-3} \text{Mpc}^{-3}$ and $\alpha = 1.76\pm0.09$).
Following on from this, applying the luminosity cut to the LM fitting procedure described in Section \ref{sec:LM} results in a slight decrease in the normalisation and slightly steeper slope for the LM relation ($A = (0.11 \pm 0.07) \times 10^{43}$\ergps and $B = 1.91 \pm 0.17$).

Applying this higher luminosity cut corresponds approximately to reducing the sample size.
While the posterior of low luminosity groups still contribute to the bins above this threshold, the sum of the posteriors integrated above this threshold is $\sim 128$, corresponding to just over half the sample.

A potential issue in constraining the luminosities of low luminosity groups is the contribution from unresolved AGN.
The 3XLSS catalogue \citep{2018AA...620A..12C_3XLSS}, used to account for point source contamination in the aperture photometry, has a 90\% completeness limit of $5.8 \times 10^{-15}$ erg s$^{-1}$ cm$^{-2}$ in the [0.5-2] keV energy band, corresponding approximately to a count rate of 1 $\times 10^{-3}$ counts s$^{-1}$.
Therefore, applying the higher luminosity cut as above ensures nearly complete knowledge of the AGN population within the fitted range, demonstrates that unresolved AGN have a negligible impact, as the results remain consistent within 1 $\sigma$.

We also need to consider the contributions of X-ray binaries within galaxies.
Studies of the the luminosity functions of low-mass and high-mass X-ray binaries (LMXBs and HMXBs, respectively) typically find a cut-off luminosity of $\sim 5 \times 10^{40}$ \ergps \citep{2019ApJS..243....3L, 2022MNRAS.512.3284S}.
These works extend to higher redshifts and cover broader energy bands than the GAMA-XXL sample, and so the flux contribution from even the brightest HMXB is likely to be negligible above the luminosity cut applied to this work.

The strength of the method used in this paper is that the X-ray luminosity contribution of all groups is included in order to analyse a complete optically selected sample.
The luminosity cut used to obtain the main results of this paper ($10^{40.6}$ \ergps) appears to adequately account for the limitation in the aperture photometry.
Applying a more conservative X-ray luminosity lower limit to account for the limitations in recovering the luminosity of low count rate groups has no meaningful impact on our conclusions.

\subsection{Comparison to other studies}
\label{sec:discuss_LM}

\begin{figure*}
    \centerline{\includegraphics[width=1.6\columnwidth]{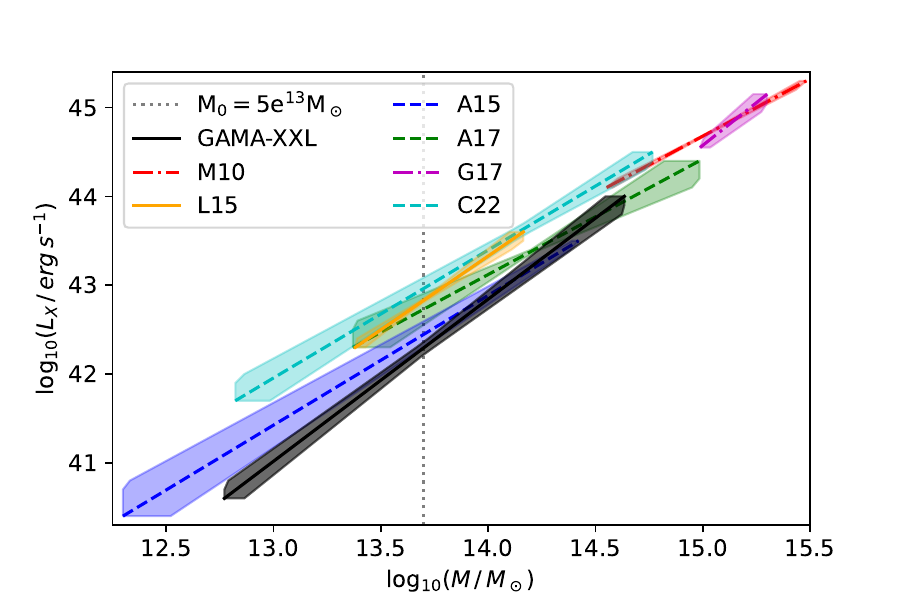} }
    \caption{Visual representation of equation \ref{eq:LM} comparing the best fitting parameters from Figure \ref{fig:LMcorner} to results from \citet[][\Mantz]{Mantz2010TheRelations}, \citet[][\Lovisari]{ 2015AA...573A.118Lovisari_Scaling}, \citet[][\Anderson]{ 2015MNRAS.449.3806Anderson_scaling}$^\dagger$, \citet[][\Andreon]{Andreon2017}$^\dagger$, \citet[][\Giles]{ Giles2017ChandraRelation}, and \citet[][\Chiu]{ Chiu2022TheSurvey} ($^\dagger$ optically selected). 
    Shaded regions indicate the $1\sigma$ error, shown only within the data-fitting range for each respective sample. 
    The vertical line shows the mass pivot used in this work.}
    \label{fig:LM}
\end{figure*}

\begin{figure}
     \centerline{\includegraphics[width=1.1\columnwidth]{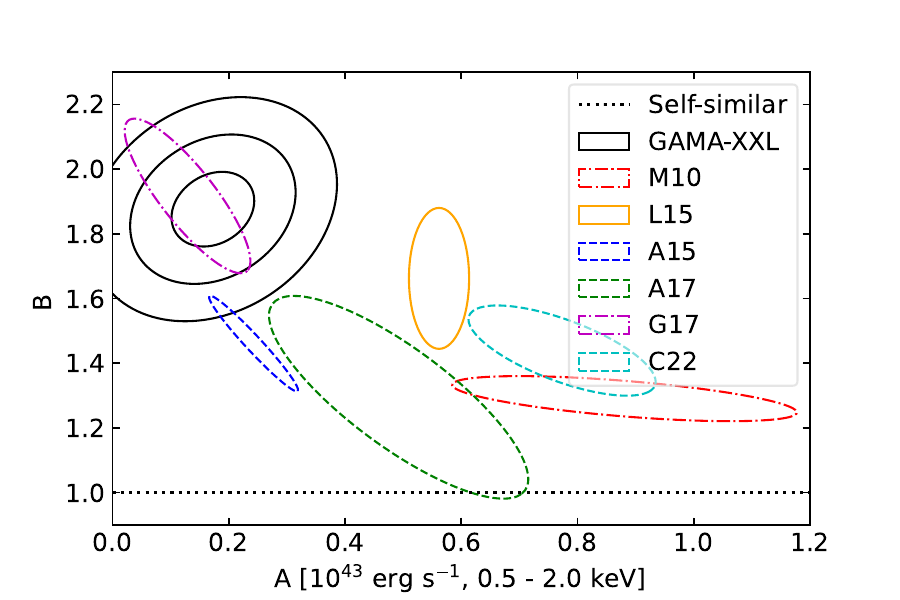} }
    \caption{Best-fitting parameters for equation \ref{eq:LM} required to reproduce the observed XLF of the GAMA groups given a theoretical HMF. 
    Contours indicate $1\sigma$, $2\sigma$, and $3\sigma$ confidence intervals and the horizontal dashed line shows the self-similar slope.  
    The slope and normalisation values of comparison works (listed in Figure \ref{fig:LM}) are converted to our pivot mass, $M_0 = 5 \times 10^{13} M_{\odot}$ (see section \ref{sec:discuss_LM} for details).
    All contours agree at $3\sigma$ or better, but only $1\sigma$ contours are shown for clarity. }
    \label{fig:LM_AB}
\end{figure}

Previous work on the LM relation has typically been confined to the study of relatively massive systems, from X-ray selected samples, and using per-cluster mass estimates allowing modelling of the LM relation directly in the LM plane. 
Measurements of the LM relation for lower mass groups are less common, due mainly to the difficulty in obtaining mass estimates for these fainter systems. 
Similarly, measurements of the LM relation for optically-selected samples, while desirable due to the minimisation of X-ray selection biases, are also rare due to the difficulties in obtaining uniform X-ray data and mass estimates.
Our analysis is thus somewhat unusual in the use of uniform X-ray data, the low masses of the systems, and the use of the XLF and HMF to infer the LM relation. 
Table \ref{tab:LM_AB} summarises several recent studies with which we compare our results. 
These are chosen to include both optically-selected samples and X-ray selected samples for which the selection biases were corrected. 

\citet[][\Mantz]{Mantz2010TheRelations} performed an analysis similar to ours but limited to more massive systems. 
They simultaneously fitted scaling relations and cosmological parameters by directly mapping the observed XLF onto a theoretical HMF for an X-ray selected sample.
An updated analysis using a similar method was conducted by \citet{Mantz2016WeighingRelations} on a subsample of 139 massive clusters which have follow up \textit{Chandra} data.
They simultaneously constrain the scaling relations of cluster X-ray luminosity, temperature and gas mass with mass and redshift, using mass information from weak gravitational lensing for 27 of the 139 clusters.
The two studies show a good agreement in the results for the slope and normalisation of the LM relation, and as such any discussion of \Mantz{} can be extended to \citet{Mantz2016WeighingRelations}.
\citet[][\Lovisari]{2015AA...573A.118Lovisari_Scaling} analysed \textit{XMM-Newton} observations for a sample of 20 RASS selected low mass groups, and constrained the LM relation using hydrostatic mass estimates for each cluster.
\citet[][\Anderson]{2015MNRAS.449.3806Anderson_scaling} performed a similar analysis to ours, but derived the XLF by stacking X-ray data from brightest cluster galaxies, enabling them to analyse a large number of groups down to lower luminosities and masses than our sample.
\citet[][\Andreon]{Andreon2017} use a sample of 34 galaxy clusters selected from the C4 catalogue \citep{Miller2005TheSurvey} using SDSS data \citep{Abazajian2004TheSurvey}.
They used archival \textit{XMM-Newton}, \textit{Chandra}, and \textit{Swift} observations alongside caustic mass analysis of the SDSS data to constrain the LM relation of an X-ray independent sample.
\citet[][\Giles]{Giles2017ChandraRelation} studied a sample of massive X-ray selected clusters and constrained the LM relation using masses determined for each cluster from a hydrostatic analysis. 
They used the HMF (assuming a fixed cosmology) to predict the expected number of clusters in order to correct for X-ray selection bias.
\citet[][\Chiu]{Chiu2022TheSurvey} use a sample of 313 optically confirmed groups from the eFEDS which are also uniformly covered by the HSC survey.
This allows the modelling of the X-ray luminosity as a function of cluster mass and redshift, jointly with the modelling of the count rate and weak-lensing mass calibration.

Figure \ref{fig:LM_AB} shows the normalisation $A$ and slope $B$ of the LM relation measured by these different studies.
In general, normalisations were measured at different pivot masses ($M_0$) for each study and so are corrected to match our $M_0=5\times10^{13}M_\odot$.
This correction is dependent on the slope of the LM relation in question and so introduces a degeneracy between $A$ and $B$ which is apparent in Figure \ref{fig:LM_AB}. 
This conversion was done by sampling pairs of $A$ and $B$ at the original pivot mass, assuming they are independent and using the reported errors. 
Each $A$, $B$ pair was then used to convert $A$ to our choice of pivot. 
In the work by \Mantz{}, \Lovisari{}, and \Giles{}, luminosities were measured in the $0.1-2.4$ keV energy band, while the work by \Anderson{} uses bolometric luminosities.
For each sample, we calculate a correction factor to convert the luminosities to the 0.5-2.0 keV energy band at the temperature determined by that sample's mass pivot, using an APEC model and the MT relation by \citet{Umetsu2020} as described in Section \ref{sec:Photometry_discussion}.
The mass pivots for each sample, along with the calculated temperature and any energy band corrections are detailed in Table \ref{tab:LM_AB}. 

In Figure \ref{fig:LM_AB} we only plot the $1\sigma$ contours for the literature values but in all cases the 3$\sigma$ contours overlap.
However, it is clear that there is not a strong consensus and much of the parameter space is covered by the different studies. 
The large uncertainty in the normalisations of some LM relations shown in Figure \ref{fig:LM_AB} (e.g., \Mantz, \Andreon) is due to the uncertainty on the slope and their larger pivot mass, and reflects the uncertainty in extrapolating high-mass samples to our low-mass regime.
In fact, the normalisations of the \Mantz{} and \Giles{} relations agree well in the mass ranges covered by their data ($10^{14.5}-10^{15}$ M$_\odot$), and the difference in Figure \ref{fig:LM_AB} is due to the differences in slope. 
This can be seen more clearly in Figure \ref{fig:LM}, which shows a visual representation of the LM relation across the luminosity and mass range explored by each study.

Our analysis strongly favours a slope of the LM relation in the low-mass regime that is steeper than the self-similar expectation of $B \approx 1$. 
This is somewhat consistent with the results of \Lovisari{}, \Anderson{}, and \Andreon{} on low-mass systems. 
In comparison to the studies of high-mass clusters, our slope is similar to that found by \Giles, but somewhat steeper than the slope found by \Mantz. 
Despite the tension between \Giles{} and other X-ray selected samples of high mass clusters, we see a trend in decreasing slope with increasing mass regime.
It is important to note that the self-similar slope of $B \approx 1$ is based on the assumption that the X-ray emission is solely due to bremsstrahlung. 
For the lower mass systems in our sample, line emission makes a significant contribution to the emission, which would flatten the self-similar relation to $B < 1$ at low masses. 
Thus, the difference we find from self-similarity is a lower limit.

The normalisation of the LM relation at $M_0 = 5 \times 10^{13} M_{\odot}$ also shows considerable variation between studies. 
In comparison with the other low-mass samples, we find good agreement with the optically-selected sample of \Anderson, but a lower normalisation than the X-ray selected samples of \Lovisari{} and \Chiu. 
This could be a consequence of the X-ray selection being biased towards the X-ray bright portion of the cluster population at a given mass. 
We note that \Lovisari{} made an approximate correction for this bias using simulated data.
However, their approach assumed a scatter in luminosity that was measured without correcting for the bias. 
If the real scatter were larger, then the bias would be underestimated, which could explain the difference in normalisation.
The LM relation found by \Chiu{} also aims to account for the missing low luminosity groups by using simulations.
However, as described in \citet{Liu2022TheGroups}, the X-ray images of groups below $5 \times 10^{13} M_\odot$ are simulated by reducing the flux of higher mass groups, potentially leading to groups that are too bright.
This could potentially explain some of the differences in the low mass regime.
While the normalisation's found by \Andreon{} and \Giles{} may appear to be in contradiction to these findings, this may be an artificial impact due to the degeneracy between the slope and normalisation of the LM relation when correcting for the change in mass pivot.
The normalisation of the optically selected sample by \Andreon{} is lower than X-ray selected samples when considered at the mass pivot used to fit the LM relation (given in Table \ref{tab:LM_AB}), and the inverse can be seen for \Giles.
Overall, the discrepancies between X-ray and optically selected clusters compared in this paper are consistent with claims that X-ray selected samples may be missing some population of clusters, and that this is not fully accounted for \citep{Andreon2022LowRelations}. 
Figure \ref{fig:LM} illustrates that, when looking at the combined picture of three optically selected samples, this issue is apparent from the smallest groups in \Anderson{} to the more massive clusters in \Andreon.
The results also highlight the large scatter in X-ray luminosity for a given mass, which was not directly measured in this work.

The purpose of our work was to explore the X-ray properties of an optically selected sample of galaxy groups in order to avoid the impact of X-ray selection effects.
Similar work by \PaperXLV{} studied a sub-sample of 142 of the GAMA-XXL sample by comparing the X-ray and optical properties of the groups.
The sample was divided into categories with low, median and high X-ray luminosities relative to their optical luminosities.
It would be expected that X-ray overluminous groups would be more dynamically evolved when compared to X-ray underluminous groups, as there would have been more time for the ICM to become virialised and emit strongly in the X-ray.
\PaperXLV{} found that most X-ray overluminous groups were more dynamically evolved than underluminous groups.
However, 20\% of X-ray underluminous groups were found to be dynamically evolved, most likely due to high gas entropy resulting from feedback mechanisms.

X-ray underluminous groups would be strongly underrepresented in X-ray selected samples, and due to the lack of coverage of high quality X-ray data in the sky, even optically selected samples would have little X-ray data for such work.
The benefit of our method is the inclusion of these groups, and due to our approach using forced X-ray aperture photometry, we are able to include X-ray underluminous groups in our analysis.
This population of X-Ray underluminous groups may account for some of the differences found in the normalisation of the LM relation between optical and X-ray selected samples.
As such, including X-ray underluminous groups is an alternative method to attempting to correct for X-ray selection. 
This approach is not without selection biases which need understanding, such as how optical contamination or line-of-sight elongations contribute to the low luminosity population.
Both methods can be refined and compared to build up a more comprehensive understanding of the low mass regime where selection effects dominate.
Hopefully larger complete samples from Euclid observed in X-ray by eROSITA will help to solve these discrepancies.

\subsection{Astrophysical Implications}
\label{sec:comp_to_sims}

Extrapolating the LM relation of various studies (as shown in Figure \ref{fig:LM}) shows that the LM relation is observationally not well constrained at lower (and higher) masses.
However, most of these results can be reconciled by a broken power-law which steepens at masses of around $10^{14} M_\odot$ (\Lovisari).
This combines the steeper slope seen in lower mass regimes with the shallower (near self-similar) slope seen for the highest mass objects.
Due to the difficulty in observing complete samples over a large enough mass range, constraints on a broken power-law fit for the LM relation has been limited to simulated data.

\begin{figure}
    \centerline{\includegraphics[width=1.1\columnwidth]{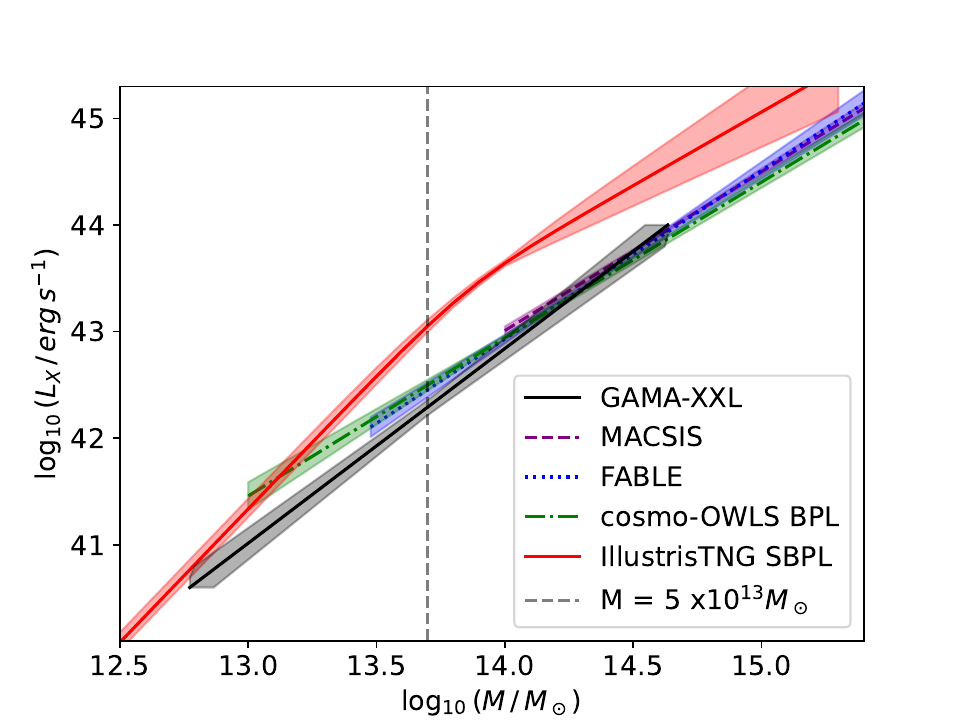} }
    \caption{Visual representation of equation \ref{eq:LM} as described in Figure \ref{fig:LM} with comparison to simulations such as MACSIS \citep{Barnes2017TheSimulations}, FABLE \citep{Henden2019TheSimulations}, cosmo-OWLS \citep[][broken power-law]{Brun2017ThePopulations}, and IllustrisTNG \citep[][smoothly broken powerlaw]{Pop2022Sunyaev-ZeldovichSimulations}.
    }
    \label{fig:LM_sims}
\end{figure}

Most simulations which investigate the LM relation fit a power-law model such as our own, and find slopes that are steeper than self-similar: $1.58 \pm 0.1$ in FABLE \citep{Henden2019TheSimulations}, $1.49 \pm 0.05$ in MACSIS \citep{Barnes2017TheSimulations}, and $1.422 \pm 0.019$ in cosmo-OWLS \citep{Brun2017ThePopulations}\footnote{The slope is bandpass dependent and quoted results have been corrected from bolometric luminosity to the 0.5-2.0 keV energy band used in this work by correcting the slope downwards by 0.39 and the normalisation by a factor of 0.23.}.
These slopes are slightly shallower than our results, likely due to the higher mass ranges used, but follow the observed trend in Figure \ref{fig:LM}.
\citet{Brun2017ThePopulations} also fit a broken power-law (BPL) with a fixed pivot at $10^{14} M_\odot$, finding that the low mass regime was best described by a slope of $1.49 \pm 0.1$, which is still shallower than our slope, while extending down to masses of $1 \times 10^{13} M_{\odot}$.
The high mass regime slope ($1.456 \pm 0.036$) is similar to that found using observations of high mass clusters.
\citet{Pop2022Sunyaev-ZeldovichSimulations} followed this up using the IllustrisTNG simulation, extending down to masses of $1 \times 10^{12} M_{\odot}$.
They found that the scaling relations are best described using a smoothly broken power-law (SBPL), with breaks occurring in the mass range $ 0.3 - 2 \times 10^{14} M_\odot$.
The slope for the low mass regime ($2.503 \pm 0.009$) is much steeper than that found in our work, despite agreement with other work for the high mass regime ($1.379 \pm 0.272$).

A comparison of the LM relations found using simulations is shown in Figure \ref{fig:LM_sims}.
With the notable exception of IllustrisTNG, the scaling relations found using simulations are in closer agreement with the optically selected GAMA-XXL sample than to X-ray selected samples (Figure \ref{fig:LM}).
This is to be expected given that the samples used in the simulations were selected using a FoF algorithm, similar to that used by the GAMA-XXL sample which was calibrated using simulations (as described in \citet{GAMA}).
\citet{Pop2022Sunyaev-ZeldovichSimulations} observe that galaxy groups in the mass range $ 0.5 - 2 \times 10^{14} M_\odot$ appear to be too gas rich in IllustrisTNG, which may account for the significantly higher normalisation found in this range.

The predictions of simulations are highly sensitive to the subgrid modelling used to calibrate the astrophysical processes which originate on unresolved scales.
As such, detailed observational constraints are required with which the results of simulations can be fine tuned and compared.
\citet{Brun2017ThePopulations} uses a build-up thermal injection model following the prescription by \citet{Booth2009CosmologicalTests}, where black holes accumulate the feedback energy, at a rate of 1.5\% of the rest-mass energy of the gas, until they can heat neighbouring gas particles by $10^{8.5}$ K.
\citet{Pop2022Sunyaev-ZeldovichSimulations} on the other hand uses a dual AGN model \citep[described in][]{Weinberger2018SupermassiveSimulation}, where thermal energy is continuously injected into the surrounding, while kinetic energy is built up and injected each time the black hole grows by 15\%.
This AGN feedback model is less effective at removing gas from the halo compared to the model used in earlier Illustris simulations, potentially resulting in gas rich halos.
Determining which simulation shows best agreement with our work is not possible due to the different parameters fit and the different mass ranges used.

\section{Summary and Conclusions}
\label{sec:conclusions}

The overlap between the GAMA optical survey and the XXL X-ray survey was used to explore the X-ray properties of an optically selected sample of 235 galaxy groups.
Forced X-ray photometry on these groups was used to measure the luminosity posterior distribution for each group.
This allowed us to measure the XLF to luminosities over an order of magnitude fainter than X-ray selected samples, and the sample is also free from X-ray selection biases.
The XLF was found to be in agreement with the extrapolation of the XLF from X-ray selected samples at higher luminosities, extending down over an order of magnitude lower.

The measured XLF was then used with a theoretical HMF to infer the form of the scaling relation between X-ray luminosity and mass at lower luminosities than previously possible from X-ray selected samples, yielding some of the most precise measurements of the LM relation in the low-mass regime. 
We find the slope of the LM relation to be $1.87 \pm 0.12$. 
Like many other studies, this is significantly steeper than the self-similar expectation, especially taking into account the flattening of the self-similar relation due to line emission in lower mass systems.
Measurements of the normalisation of the LM relation show significant variation between studies due to a combination of the mass ranges probed and the selection methods used. 

Our approach can be extended to the eROSITA all sky survey \citep{Predehl2021TheSRG} data and combined with future all-sky optical and near infra-red surveys such as the Vera Rubin Observatory \citep{Ivezic2019LSST:Products} and Euclid \citep{Laureijs2011EuclidReport}. 
The large resulting datasets will enable us to constrain more complex parametrisations of the LM relation to gain further insight into the astrophysical processes shaping the relation beyond the limits in mass and redshift accessible to X-ray selected samples.

\section*{Acknowledgements}

CW acknowledges support from STFC grants S100255-101 and ST/Y002008/1.
BJM acknowledges support from STFC grants ST/V000454/1 and ST/Y002008/1.
MP acknowledges long-term support from the Centre National d'Etudes Spatiales.
Thanks to Lucio Chiappetti for the examination of the manuscript.
We thank the anonymous referees for their comments and questions, which improved the analysis and discussion in this manuscript.
GAMA is a joint European-Australasian project based around a spectroscopic campaign using the Anglo-Australian Telescope. 
The GAMA input catalogue is based on data taken from the Sloan Digital Sky Survey and the UKIRT Infrared Deep Sky Survey. 
Complementary imaging of the GAMA regions is being obtained by a number of independent survey programmes including GALEX MIS, VST KiDS, VISTA VIKING, WISE, Herschel-ATLAS, GMRT and ASKAP providing UV to radio coverage. 
GAMA is funded by the STFC (UK), the ARC (Australia), the AAO, and the participating institutions. The GAMA website is \url{http://www.gama-survey.org/}.
Based on observations obtained with XMM-Newton, an ESA science mission with instruments and contributions directly funded by ESA Member States and NASA. 
XXL is an international project based around an XMM Very Large Programme surveying two 25 deg$^2$ extragalactic fields at a depth of $\sim 6 \times  10^{-15} $erg cm$^{-2}$ s$^{-1}$ in the [0.5-2] keV band for point-like sources. 
The XXL website is \url{http://irfu.cea.fr/xxl}.
We gratefully acknowledge the use of open software, including Matplotlib \citep{2007CSE.....9...90H}, SciPy \citep{2020NatMe..17..261V}, NumPy \citep{2020Natur.585..357H}, Astropy \citep{2013AA...558A..33A}, Topcat \citep{2005ASPC..347...29T}, DS9 \citep{Hanisch2001AstronomicalX}, and those directly cited in the paper.

\section*{Data Availability}
The data underlying this study will be shared on reasonable request to the corresponding author.




\bibliographystyle{mnras}
\bibliography{abbreferences} 




\onecolumn

\appendix
\setcounter{table}{0}
\renewcommand{\thetable}{A\arabic{table}}

\section{Energy band corrections}
\label{appendix:bandcorr}
\begin{longtable}[c]{lllllll}
    \captionsetup{width=440pt}
    \caption[width=20cm]{Details of comparative LM relation studies by \citet[][\Mantz]{Mantz2010TheRelations}, \citet[][\Lovisari]{ 2015AA...573A.118Lovisari_Scaling}, \citet[][\Anderson]{ 2015MNRAS.449.3806Anderson_scaling}$^\dagger$, \citet[][\Andreon]{Andreon2017}$^\dagger$, \citet[][\Giles]{ Giles2017ChandraRelation}, and \citet[][\Chiu]{ Chiu2022TheSurvey} ($^\dagger$ optically selected).
    The temperature for each study was calculated at the mass pivot using the MT relation by \citet{Umetsu2020}, and the conversion from the studies energy band to our 0.5-2.0 keV energy band calculated using an APEC model ($\sim 0.6$ for 0.1-2.4 keV energy band and $\sim 0.23$ for bolometric).
    }
    \label{tab:LM_AB}\\
         Paper & Sample & $L_{0.5-2.0keV}$ \ergps & Instrument & $M_0 (M_\odot)$ & T (keV) & conversion \\
         \hline
         This Work$^\dagger$ & 235 & $\gtrsim 4 \times 10^{40}$ & \textit{XMM-Newton} & $5\times 10^{13}$ & 1.3 & - \\ 
         \Mantz & 238 & $\gtrsim 1.5 \times 10^{44}$ & \textit{ROSAT}/ \textit{Chandra}  & $1\times 10^{15}$ & 16.8 & 0.62 \\
         \Lovisari & 20 & $0.2-3.2 \times 10^{43}$ & \textit{XMM-Newton} & $5\times 10^{13}$ & 1.3 & 0.65 \\ 
         \Anderson$^\dagger$ & $\sim$12k & $\gtrsim 2.5 \times 10^{40}$ & \textit{ROSAT} & $4\times 10^{14}$ & 7.7 & 0.23  \\
         \Andreon$^\dagger$ & 34 & $\gtrsim 2 \times 10^{42}$ & \textit{Swift}/ \textit{Chandra}/ \textit{XMM} & $1.6\times 10^{14}$ & 3.5 & - \\
         \Giles & 34 & $\gtrsim 3.6 \times 10^{44}$ & \textit{Chandra}  & $1\times 10^{15}$ & 16.8 & 0.62\\ 
         \Chiu & 313 & $\gtrsim 0.5 \times 10^{42}$ & \textit{eROSITA}  & $1.4\times 10^{14}$ & 3.2 & - \\ 
    
\end{longtable}
\section{Luminosities}
\label{appendix:LumRes}

\renewcommand{\thetable}{B\arabic{table}}

\begin{longtable}[c]{@{}ccccccccccc@{}}
\captionsetup{width=440pt}
\caption[width=20cm]{X-ray luminosities of the optically selected GAMA-XXL group sample given in units \ergps in the 0.5-2.0 keV energy band. GroupID is unique to the GAMA \citep{2011MNRAS.413..971Driver_GAMA} group catalogue DMU version G3Cv10 obtained from \url{http://www.gama-survey.org/dr3/data/cat/GroupFinding/v10/}. Also given is each groups position, richness ($N_\text{FoF}$), dynamical mass (from \cite{GAMA}) and core radius ($r_c$).
}
\label{tab:Luminosities}\\
\toprule
   GroupID &RA &DEC &\textit{z} & N\textsubscript{FoF} & L\textsubscript{X, Mode} & L\textsubscript{X, Median} & L\textsubscript{X, 16\%} & L\textsubscript{X, 84\%} & Mass &     r\textsubscript{c} \\
    &$^\circ$ &$^\circ$ & & & \ergps & \ergps & \ergps & \ergps & $M_\odot$ & kpc \\* \midrule
\endfirsthead
\multicolumn{11}{c}
{{\bfseries Table \thetable\ continued from previous page}} \\
\toprule
   GroupID &RA &DEC &\textit{z} & N\textsubscript{FoF} & L\textsubscript{X} Mode & L\textsubscript{X} Median & L\textsubscript{X} 16\% & L\textsubscript{X} 84\% & Mass &     r\textsubscript{c} \\* \midrule
\endhead
\bottomrule
\endfoot
\endlastfoot
    400001 & 33.671 & -4.567 & 0.138 &       97 &    9.79$\times10^{43}$ &      9.79$\times10^{43}$ &   9.72$\times10^{43}$ & 9.86$\times10^{43}$ &    4.12$\times10^{14}$ & 156 \\
    400002 & 37.922 & -4.883 & 0.184 &       59 &    1.56$\times10^{44}$ &      1.56$\times10^{44}$ &   1.54$\times10^{44}$ & 1.58$\times10^{44}$ &    6.76$\times10^{14}$ & 181 \\
    400003 & 36.319 & -5.887 & 0.053 &       54 &    3.03$\times10^{42}$ &      3.04$\times10^{42}$ &   2.85$\times10^{42}$ & 3.22$\times10^{42}$ &    1.25$\times10^{14}$ & 107 \\
    400004 & 33.112 & -5.627 & 0.299 &       41 &    7.92$\times10^{43}$ &      8.69$\times10^{43}$ &   8.33$\times10^{43}$ & 9.06$\times10^{43}$ &    8.96$\times10^{14}$ & 191 \\
    400007 & 36.455 & -5.896 & 0.232 &       19 &    2.59$\times10^{43}$ &      2.59$\times10^{43}$ &   2.48$\times10^{43}$ & 2.70$\times10^{43}$ &    3.79$\times10^{14}$ & 146 \\
    400008 & 31.350 & -5.737 & 0.301 &       23 &    3.27$\times10^{43}$ &      3.27$\times10^{43}$ &   3.09$\times10^{43}$ & 3.45$\times10^{43}$ &    4.10$\times10^{14}$ & 147 \\
    400009 & 30.496 & -4.075 & 0.166 &       17 &    1.52$\times10^{42}$ &      1.52$\times10^{42}$ &   1.01$\times10^{42}$ & 2.04$\times10^{42}$ &    6.24$\times10^{13}$ &  82 \\
    400010 & 37.662 & -4.991 & 0.291 &       16 &    7.11$\times10^{42}$ &      2.44$\times10^{43}$ &   2.26$\times10^{43}$ & 2.61$\times10^{43}$ &    3.84$\times10^{14}$ & 144 \\
    400012 & 32.762 & -4.894 & 0.137 &       23 &    1.68$\times10^{43}$ &      1.68$\times10^{43}$ &   1.61$\times10^{43}$ & 1.75$\times10^{43}$ &    2.02$\times10^{14}$ & 123 \\
    400013 & 34.020 & -4.233 & 0.153 &       20 &    1.21$\times10^{43}$ &      1.21$\times10^{43}$ &   1.17$\times10^{43}$ & 1.26$\times10^{43}$ &    1.29$\times10^{14}$ & 105 \\
    400014 & 30.977 & -4.232 & 0.137 &       29 &    1.31$\times10^{42}$ &      1.30$\times10^{42}$ &   1.05$\times10^{42}$ & 1.56$\times10^{42}$ &    1.07$\times10^{14}$ &  99 \\
    400015 & 32.509 & -5.468 & 0.137 &       17 &    --- &      2.03$\times10^{41}$ &   6.17$\times10^{40}$ & 4.21$\times10^{41}$ &    7.34$\times10^{13}$ &  88 \\
    400016 & 36.487 & -5.643 & 0.194 &       15 &    4.87$\times10^{42}$ &      4.87$\times10^{42}$ &   4.43$\times10^{42}$ & 5.31$\times10^{42}$ &    1.45$\times10^{14}$ & 108 \\
    400019 & 37.891 & -5.262 & 0.141 &       19 &    5.07$\times10^{40}$ &      1.34$\times10^{41}$ &   4.22$\times10^{40}$ & 2.67$\times10^{41}$ &    2.55$\times10^{13}$ &  61 \\
    400020 & 35.485 & -5.759 & 0.258 &       14 &    1.00$\times10^{43}$ &      1.00$\times10^{43}$ &   9.32$\times10^{42}$ & 1.07$\times10^{43}$ &    8.50$\times10^{13}$ &  88 \\
    400021 & 33.873 & -5.549 & 0.289 &       13 &    9.24$\times10^{42}$ &      9.25$\times10^{42}$ &   7.98$\times10^{42}$ & 1.05$\times10^{43}$ &    6.45$\times10^{14}$ & 171 \\
    400023 & 37.624 & -5.225 & 0.187 &       12 &    5.18$\times10^{42}$ &      5.18$\times10^{42}$ &   4.66$\times10^{42}$ & 5.71$\times10^{42}$ &    8.12$\times10^{13}$ &  89 \\
    400025 & 32.459 & -4.369 & 0.138 &       19 &    2.03$\times10^{42}$ &      2.03$\times10^{42}$ &   1.77$\times10^{42}$ & 2.29$\times10^{42}$ &    4.19$\times10^{13}$ &  73 \\
    400026 & 36.573 & -5.079 & 0.054 &       42 &    1.44$\times10^{41}$ &      1.94$\times10^{42}$ &   1.83$\times10^{42}$ & 2.05$\times10^{42}$ &    1.07$\times10^{14}$ & 102 \\
    400027 & 31.669 & -5.344 & 0.137 &       11 &    2.08$\times10^{42}$ &      2.09$\times10^{42}$ &   1.69$\times10^{42}$ & 2.48$\times10^{42}$ &    1.48$\times10^{14}$ & 111 \\
    400028 & 31.618 & -4.686 & 0.135 &       11 &    9.54$\times10^{41}$ &      9.53$\times10^{41}$ &   7.73$\times10^{41}$ & 1.13$\times10^{42}$ &    3.05$\times10^{13}$ &  65 \\
    400029 & 30.817 & -5.405 & 0.210 &       12 &    1.71$\times10^{42}$ &      1.73$\times10^{42}$ &   9.83$\times10^{41}$ & 2.48$\times10^{42}$ &    6.51$\times10^{14}$ & 177 \\
    400030 & 34.983 & -5.468 & 0.277 &       13 &    --- &      3.31$\times10^{41}$ &   1.00$\times10^{41}$ & 6.94$\times10^{41}$ &    1.30$\times10^{14}$ & 101 \\
    400031 & 36.476 & -5.341 & 0.053 &       16 &    1.10$\times10^{41}$ &      1.10$\times10^{41}$ &   9.75$\times10^{40}$ & 1.22$\times10^{41}$ &    1.13$\times10^{13}$ &  48 \\
    400032 & 36.618 & -4.014 & 0.208 &       12 &    6.51$\times10^{41}$ &      6.52$\times10^{41}$ &   3.88$\times10^{41}$ & 9.20$\times10^{41}$ &    8.66$\times10^{13}$ &  90 \\
    400033 & 33.915 & -5.980 & 0.307 &       10 &    1.26$\times10^{43}$ &      1.26$\times10^{43}$ &   1.15$\times10^{43}$ & 1.38$\times10^{43}$ &    3.89$\times10^{14}$ & 144 \\
    400034 & 33.800 & -4.748 & 0.292 &       10 &    3.93$\times10^{41}$ &      7.05$\times10^{41}$ &   2.33$\times10^{41}$ & 1.36$\times10^{42}$ &    1.67$\times10^{14}$ & 109 \\
    400035 & 32.121 & -4.942 & 0.306 &       11 &    6.36$\times10^{42}$ &      6.36$\times10^{42}$ &   5.32$\times10^{42}$ & 7.39$\times10^{42}$ &    2.45$\times10^{14}$ & 123 \\
    400036 & 30.590 & -5.903 & 0.234 &       14 &    2.66$\times10^{42}$ &      2.66$\times10^{42}$ &   1.96$\times10^{42}$ & 3.35$\times10^{42}$ &    1.34$\times10^{14}$ & 103 \\
    400037 & 35.367 & -4.570 & 0.158 &       10 &    2.13$\times10^{41}$ &      2.13$\times10^{41}$ &   1.32$\times10^{41}$ & 2.94$\times10^{41}$ &    3.40$\times10^{13}$ &  67 \\
    400039 & 36.813 & -5.076 & 0.142 &       17 &    6.99$\times10^{40}$ &      9.47$\times10^{40}$ &   3.36$\times10^{40}$ & 1.73$\times10^{41}$ &    5.43$\times10^{13}$ &  79 \\
    400040 & 38.138 & -4.775 & 0.276 &       12 &    7.25$\times10^{42}$ &      7.26$\times10^{42}$ &   6.13$\times10^{42}$ & 8.38$\times10^{42}$ &    3.79$\times10^{14}$ & 144 \\
    400041 & 36.864 & -4.650 & 0.070 &       32 &    1.56$\times10^{42}$ &      1.56$\times10^{42}$ &   1.48$\times10^{42}$ & 1.65$\times10^{42}$ &    1.02$\times10^{14}$ & 100 \\
    400042 & 38.414 & -4.175 & 0.275 &       11 &    6.18$\times10^{42}$ &      6.17$\times10^{42}$ &   5.39$\times10^{42}$ & 6.95$\times10^{42}$ &    2.79$\times10^{13}$ &  60 \\
    400043 & 34.219 & -4.291 & 0.152 &       12 &    1.44$\times10^{41}$ &      2.96$\times10^{41}$ &   9.58$\times10^{40}$ & 5.80$\times10^{41}$ &    1.02$\times10^{14}$ &  97 \\
    400045 & 33.868 & -4.678 & 0.347 &       11 &    6.67$\times10^{43}$ &      6.67$\times10^{43}$ &   6.35$\times10^{43}$ & 6.98$\times10^{43}$ &    1.02E+15 & 196 \\
    400046 & 32.756 & -5.062 & 0.070 &       13 &    9.90$\times10^{40}$ &      1.17$\times10^{41}$ &   4.49$\times10^{40}$ & 2.06$\times10^{41}$ &    4.86$\times10^{13}$ &  78 \\
    400047 & 31.834 & -4.825 & 0.253 &       10 &    1.67$\times10^{43}$ &      1.67$\times10^{43}$ &   1.56$\times10^{43}$ & 1.77$\times10^{43}$ &    1.02$\times10^{14}$ &  94 \\
    400048 & 31.214 & -4.662 & 0.106 &       16 &    --- &      4.89$\times10^{40}$ &   1.46$\times10^{40}$ & 1.02$\times10^{41}$ &    1.18$\times10^{13}$ &  48 \\
    400049 & 36.138 & -4.239 & 0.263 &       11 &    1.92$\times10^{42}$ &      1.92$\times10^{42}$ &   1.69$\times10^{42}$ & 2.16$\times10^{42}$ &    5.38$\times10^{13}$ &  76 \\
    400050 & 37.329 & -5.887 & 0.291 &       10 &    2.36$\times10^{43}$ &      2.54$\times10^{43}$ &   2.31$\times10^{43}$ & 2.78$\times10^{43}$ &    4.69$\times10^{14}$ & 154 \\
    400051 & 38.633 & -4.932 & 0.139 &       11 &    2.04$\times10^{42}$ &      2.04$\times10^{42}$ &   1.75$\times10^{42}$ & 2.34$\times10^{42}$ &    4.03$\times10^{13}$ &  72 \\
    400053 & 36.557 & -5.446 & 0.243 &       10 &    3.59$\times10^{40}$ &      2.16$\times10^{41}$ &   6.60$\times10^{40}$ & 4.42$\times10^{41}$ &    2.17$\times10^{14}$ & 121 \\
    400054 & 36.234 & -5.133 & 0.083 &        9 &    8.11$\times10^{41}$ &      8.11$\times10^{41}$ &   7.79$\times10^{41}$ & 8.44$\times10^{41}$ &    2.20$\times10^{13}$ &  60 \\
    400055 & 37.455 & -4.090 & 0.257 &       14 &    4.04$\times10^{41}$ &      5.59$\times10^{41}$ &   1.97$\times10^{41}$ & 1.02$\times10^{42}$ &    2.46$\times10^{14}$ & 126 \\
    400056 & 34.133 & -4.508 & 0.446 &        9 &    2.32$\times10^{43}$ &      5.58$\times10^{43}$ &   5.20$\times10^{43}$ & 5.96$\times10^{43}$ &    3.16E+15 & 275 \\
    400059 & 33.770 & -5.970 & 0.254 &        8 &    8.02$\times10^{42}$ &      8.02$\times10^{42}$ &   7.20$\times10^{42}$ & 8.85$\times10^{42}$ &    1.85$\times10^{14}$ & 115 \\
    400060 & 32.463 & -5.673 & 0.137 &        8 &    1.25$\times10^{42}$ &      1.25$\times10^{42}$ &   9.96$\times10^{41}$ & 1.51$\times10^{42}$ &    2.64$\times10^{13}$ &  62 \\
    400061 & 32.440 & -5.325 & 0.137 &       10 &    --- &      9.85$\times10^{40}$ &   2.96$\times10^{40}$ & 2.06$\times10^{41}$ &    3.87$\times10^{13}$ &  71 \\
    400062 & 33.018 & -4.763 & 0.070 &        9 &    --- &      3.16$\times10^{40}$ &   9.40$\times10^{39}$ & 6.58$\times10^{40}$ &    1.79$\times10^{13}$ &  56 \\
    400064 & 31.024 & -5.548 & 0.075 &        8 &    1.99$\times10^{40}$ &      6.91$\times10^{40}$ &   2.15$\times10^{40}$ & 1.40$\times10^{41}$ &    2.54$\times10^{13}$ &  63 \\
    400065 & 30.421 & -5.030 & 0.232 &       10 &    4.95$\times10^{42}$ &      4.94$\times10^{42}$ &   3.89$\times10^{42}$ & 5.99$\times10^{42}$ &    1.42$\times10^{14}$ & 105 \\
    400066 & 30.665 & -4.150 & 0.042 &        9 &    2.65$\times10^{41}$ &      2.65$\times10^{41}$ &   2.42$\times10^{41}$ & 2.88$\times10^{41}$ &    2.64$\times10^{12}$ &  30 \\
    400068 & 36.754 & -5.663 & 0.316 &        8 &    4.33$\times10^{42}$ &      4.32$\times10^{42}$ &   3.39$\times10^{42}$ & 5.26$\times10^{42}$ &    7.33$\times10^{13}$ &  82 \\
    400069 & 36.771 & -5.104 & 0.150 &       13 &    4.37$\times10^{41}$ &      4.36$\times10^{41}$ &   3.30$\times10^{41}$ & 5.42$\times10^{41}$ &    6.56$\times10^{13}$ &  84 \\
    400070 & 36.121 & -4.165 & 0.263 &        9 &    --- &      1.25$\times10^{41}$ &   3.75$\times10^{40}$ & 2.60$\times10^{41}$ &    1.48$\times10^{14}$ & 106 \\
    400072 & 37.021 & -5.295 & 0.315 &        8 &    5.76$\times10^{42}$ &      5.76$\times10^{42}$ &   5.13$\times10^{42}$ & 6.39$\times10^{42}$ &    1.16$\times10^{14}$ &  96 \\
    400073 & 37.122 & -4.857 & 0.141 &        9 &    3.70$\times10^{42}$ &      3.70$\times10^{42}$ &   3.35$\times10^{42}$ & 4.05$\times10^{42}$ &    5.40$\times10^{13}$ &  79 \\
    400074 & 38.215 & -5.566 & 0.217 &        8 &    1.21$\times10^{42}$ &      1.21$\times10^{42}$ &   8.68$\times10^{41}$ & 1.55$\times10^{42}$ &    4.95$\times10^{13}$ &  75 \\
    400075 & 38.502 & -4.827 & 0.179 &        8 &    2.87$\times10^{42}$ &      2.87$\times10^{42}$ &   2.38$\times10^{42}$ & 3.37$\times10^{42}$ &    1.94$\times10^{14}$ & 119 \\
    400076 & 38.067 & -4.026 & 0.182 &       14 &    2.26$\times10^{42}$ &      2.26$\times10^{42}$ &   1.71$\times10^{42}$ & 2.80$\times10^{42}$ &    1.15$\times10^{14}$ & 100 \\
    400078 & 31.809 & -5.969 & 0.087 &       10 &    6.63$\times10^{41}$ &      6.63$\times10^{41}$ &   5.76$\times10^{41}$ & 7.50$\times10^{41}$ &    2.89$\times10^{13}$ &  65 \\
    400079 & 35.282 & -4.999 & 0.139 &        8 &    1.05$\times10^{42}$ &      1.05$\times10^{42}$ &   9.50$\times10^{41}$ & 1.15$\times10^{42}$ &    2.39$\times10^{14}$ & 130 \\
    400080 & 35.287 & -4.652 & 0.080 &       12 &    --- &      2.74$\times10^{40}$ &   8.16$\times10^{39}$ & 5.72$\times10^{40}$ &    7.38$\times10^{13}$ &  89 \\
    400081 & 36.294 & -4.031 & 0.172 &       11 &    2.24$\times10^{42}$ &      2.24$\times10^{42}$ &   2.01$\times10^{42}$ & 2.47$\times10^{42}$ &    6.85$\times10^{13}$ &  85 \\
    400083 & 35.020 & -5.391 & 0.084 &        7 &    --- &      1.73$\times10^{40}$ &   5.20$\times10^{39}$ & 3.62$\times10^{40}$ &    1.05$\times10^{13}$ &  47 \\
    400085 & 33.817 & -5.979 & 0.285 &        7 &    1.62$\times10^{42}$ &      1.63$\times10^{42}$ &   9.85$\times10^{41}$ & 2.27$\times10^{42}$ &    5.61$\times10^{13}$ &  76 \\
    400086 & 33.982 & -5.500 & 0.252 &        7 &    --- &      5.22$\times10^{41}$ &   1.56$\times10^{41}$ & 1.09$\times10^{42}$ &    3.54$\times10^{13}$ &  66 \\
    400087 & 33.163 & -4.833 & 0.205 &        9 &    --- &      1.66$\times10^{41}$ &   4.91$\times10^{40}$ & 3.46$\times10^{41}$ &    1.56$\times10^{13}$ &  51 \\
    400088 & 33.356 & -4.515 & 0.181 &        7 &    8.31$\times10^{40}$ &      1.75$\times10^{41}$ &   5.66$\times10^{40}$ & 3.43$\times10^{41}$ &    3.21$\times10^{13}$ &  65 \\
    400090 & 32.857 & -4.156 & 0.439 &        7 &    4.25$\times10^{42}$ &      4.43$\times10^{42}$ &   2.04$\times10^{42}$ & 7.04$\times10^{42}$ &    1.10$\times10^{14}$ &  90 \\
    400094 & 30.621 & -5.968 & 0.197 &        7 &    2.09$\times10^{40}$ &      2.90$\times10^{41}$ &   8.77$\times10^{40}$ & 5.97$\times10^{41}$ &    1.13$\times10^{13}$ &  46 \\
    400095 & 30.403 & -5.807 & 0.209 &        9 &    4.20$\times10^{42}$ &      4.19$\times10^{42}$ &   3.03$\times10^{42}$ & 5.37$\times10^{42}$ &    1.64$\times10^{14}$ & 112 \\
    400096 & 30.689 & -5.000 & 0.231 &        7 &    1.57$\times10^{42}$ &      1.57$\times10^{42}$ &   1.26$\times10^{42}$ & 1.89$\times10^{42}$ &    3.07$\times10^{13}$ &  63 \\
    400097 & 30.721 & -4.081 & 0.291 &       12 &    4.64$\times10^{42}$ &      4.65$\times10^{42}$ &   3.65$\times10^{42}$ & 5.63$\times10^{42}$ &    1.17$\times10^{14}$ &  97 \\
    400098 & 35.273 & -4.684 & 0.198 &        7 &    3.53$\times10^{42}$ &      3.53$\times10^{42}$ &   3.49$\times10^{42}$ & 3.57$\times10^{42}$ &    2.18$\times10^{13}$ &  57 \\
    400099 & 35.236 & -4.481 & 0.198 &        8 &    --- &      1.02$\times10^{41}$ &   3.07$\times10^{40}$ & 2.14$\times10^{41}$ &    4.75$\times10^{13}$ &  74 \\
    400101 & 36.887 & -5.964 & 0.231 &        8 &    1.01$\times10^{43}$ &      1.01$\times10^{43}$ &   9.11$\times10^{42}$ & 1.11$\times10^{43}$ &    4.76$\times10^{13}$ &  73 \\
    400102 & 36.858 & -4.537 & 0.307 &        9 &    1.27$\times10^{42}$ &      1.27$\times10^{42}$ &   7.95$\times10^{41}$ & 1.76$\times10^{42}$ &    1.31$\times10^{14}$ & 100 \\
    400103 & 36.656 & -4.209 & 0.209 &        8 &    3.64$\times10^{40}$ &      1.13$\times10^{41}$ &   3.52$\times10^{40}$ & 2.27$\times10^{41}$ &    1.23$\times10^{14}$ & 102 \\
    400106 & 37.036 & -5.595 & 0.190 &        9 &    5.77$\times10^{42}$ &      5.77$\times10^{42}$ &   5.34$\times10^{42}$ & 6.19$\times10^{42}$ &    2.89$\times10^{13}$ &  63 \\
    400107 & 37.346 & -5.307 & 0.195 &        7 &    1.12$\times10^{42}$ &      1.12$\times10^{42}$ &   8.34$\times10^{41}$ & 1.41$\times10^{42}$ &    4.11$\times10^{12}$ &  33 \\
    400108 & 37.003 & -4.864 & 0.297 &        8 &    1.86$\times10^{43}$ &      1.86$\times10^{43}$ &   1.77$\times10^{43}$ & 1.95$\times10^{43}$ &    3.40$\times10^{14}$ & 138 \\
    400109 & 37.616 & -4.564 & 0.294 &        9 &    1.03$\times10^{43}$ &      1.03$\times10^{43}$ &   8.96$\times10^{42}$ & 1.17$\times10^{43}$ &    2.85$\times10^{14}$ & 130 \\
    400113 & 38.389 & -5.999 & 0.182 &       10 &    3.67$\times10^{42}$ &      3.68$\times10^{42}$ &   2.94$\times10^{42}$ & 4.41$\times10^{42}$ &    1.38$\times10^{14}$ & 106 \\
    400114 & 38.646 & -5.579 & 0.222 &        9 &    2.23$\times10^{41}$ &      7.26$\times10^{41}$ &   2.26$\times10^{41}$ & 1.45$\times10^{42}$ &    5.15$\times10^{13}$ &  76 \\
    400115 & 34.502 & -4.242 & 0.034 &        8 &    5.71$\times10^{40}$ &      5.71$\times10^{40}$ &   4.89$\times10^{40}$ & 6.52$\times10^{40}$ &    1.24$\times10^{13}$ &  50 \\
    400116 & 30.780 & -5.293 & 0.133 &        7 &    3.59$\times10^{41}$ &      3.59$\times10^{41}$ &   2.48$\times10^{41}$ & 4.69$\times10^{41}$ &    2.28$\times10^{13}$ &  59 \\
    400118 & 35.744 & -4.121 & 0.291 &       15 &    1.52$\times10^{42}$ &      1.53$\times10^{42}$ &   1.12$\times10^{42}$ & 1.93$\times10^{42}$ &    4.95$\times10^{14}$ & 157 \\
    400119 & 36.353 & -4.679 & 0.263 &        7 &    2.13$\times10^{43}$ &      2.13$\times10^{43}$ &   2.06$\times10^{43}$ & 2.20$\times10^{43}$ &    2.90$\times10^{14}$ & 133 \\
    400122 & 34.864 & -5.919 & 0.091 &        6 &    2.14$\times10^{41}$ &      2.14$\times10^{41}$ &   1.70$\times10^{41}$ & 2.59$\times10^{41}$ &    2.35$\times10^{12}$ &  28 \\
    400123 & 34.788 & -5.910 & 0.148 &        6 &    5.57$\times10^{41}$ &      5.58$\times10^{41}$ &   3.53$\times10^{41}$ & 7.63$\times10^{41}$ &    2.12$\times10^{13}$ &  58 \\
    400124 & 34.655 & -5.674 & 0.280 &        6 &    1.16$\times10^{43}$ &      1.16$\times10^{43}$ &   1.06$\times10^{43}$ & 1.26$\times10^{43}$ &    2.86$\times10^{14}$ & 131 \\
    400125 & 34.121 & -5.086 & 0.143 &        6 &    --- &      4.17$\times10^{40}$ &   1.24$\times10^{40}$ & 8.71$\times10^{40}$ &    3.66$\times10^{12}$ &  32 \\
    400127 & 33.087 & -4.788 & 0.220 &        6 &    --- &      2.06$\times10^{41}$ &   6.17$\times10^{40}$ & 4.28$\times10^{41}$ &    1.52$\times10^{13}$ &  50 \\
    400128 & 33.765 & -4.406 & 0.137 &        7 &    --- &      7.84$\times10^{40}$ &   2.36$\times10^{40}$ & 1.64$\times10^{41}$ &    2.98$\times10^{13}$ &  65 \\
    400129 & 33.919 & -4.205 & 0.178 &        8 &    1.60$\times10^{41}$ &      2.27$\times10^{41}$ &   8.03$\times10^{40}$ & 4.19$\times10^{41}$ &    2.69$\times10^{13}$ &  62 \\
    400130 & 32.610 & -4.704 & 0.072 &        6 &    9.37$\times10^{40}$ &      9.40$\times10^{40}$ &   5.71$\times10^{40}$ & 1.32$\times10^{41}$ &    9.12$\times10^{12}$ &  45 \\
    400131 & 32.190 & -4.150 & 0.215 &        9 &    --- &      2.31$\times10^{41}$ &   6.95$\times10^{40}$ & 4.81$\times10^{41}$ &    6.48$\times10^{13}$ &  82 \\
    400133 & 32.063 & -5.675 & 0.298 &        6 &    2.24$\times10^{42}$ &      2.25$\times10^{42}$ &   1.47$\times10^{42}$ & 3.02$\times10^{42}$ &    5.65$\times10^{13}$ &  76 \\
    400134 & 31.918 & -5.231 & 0.137 &        9 &    2.08$\times10^{42}$ &      2.08$\times10^{42}$ &   1.73$\times10^{42}$ & 2.43$\times10^{42}$ &    8.74$\times10^{13}$ &  93 \\
    400135 & 31.278 & -4.894 & 0.209 &        6 &    5.43$\times10^{42}$ &      5.43$\times10^{42}$ &   4.86$\times10^{42}$ & 6.01$\times10^{42}$ &    8.43$\times10^{13}$ &  89 \\
    400137 & 30.447 & -4.683 & 0.179 &        6 &    3.24$\times10^{42}$ &      3.24$\times10^{42}$ &   2.40$\times10^{42}$ & 4.09$\times10^{42}$ &    8.79$\times10^{12}$ &  43 \\
    400138 & 35.834 & -5.454 & 0.210 &        6 &    1.45$\times10^{42}$ &      1.45$\times10^{42}$ &   1.28$\times10^{42}$ & 1.61$\times10^{42}$ &    1.19$\times10^{13}$ &  47 \\
    400139 & 36.335 & -5.967 & 0.313 &        7 &    5.70$\times10^{42}$ &      5.71$\times10^{42}$ &   4.11$\times10^{42}$ & 7.30$\times10^{42}$ &    6.11$\times10^{14}$ & 167 \\
    400140 & 36.606 & -5.794 & 0.317 &        6 &    --- &      4.52$\times10^{41}$ &   1.36$\times10^{41}$ & 9.42$\times10^{41}$ &    3.87$\times10^{13}$ &  66 \\
    400141 & 36.118 & -5.618 & 0.295 &        6 &    2.42$\times10^{42}$ &      2.42$\times10^{42}$ &   1.77$\times10^{42}$ & 3.08$\times10^{42}$ &    1.32$\times10^{14}$ & 101 \\
    400142 & 36.836 & -4.760 & 0.053 &       10 &    1.08$\times10^{41}$ &      1.08$\times10^{41}$ &   9.87$\times10^{40}$ & 1.17$\times10^{41}$ &    1.12$\times10^{13}$ &  48 \\
    400143 & 36.649 & -4.069 & 0.345 &        6 &    8.80$\times10^{42}$ &      8.80$\times10^{42}$ &   8.12$\times10^{42}$ & 9.48$\times10^{42}$ &    7.43$\times10^{14}$ & 176 \\
    400144 & 37.222 & -4.677 & 0.211 &        6 &    1.54$\times10^{42}$ &      1.54$\times10^{42}$ &   1.11$\times10^{42}$ & 1.97$\times10^{42}$ &    5.45$\times10^{13}$ &  77 \\
    400145 & 37.567 & -4.481 & 0.269 &        6 &    7.83$\times10^{41}$ &      8.43$\times10^{41}$ &   3.65$\times10^{41}$ & 1.39$\times10^{42}$ &    5.28$\times10^{13}$ &  75 \\
    400146 & 37.633 & -4.428 & 0.207 &        6 &    1.59$\times10^{42}$ &      1.59$\times10^{42}$ &   1.18$\times10^{42}$ & 2.00$\times10^{42}$ &    2.32$\times10^{13}$ &  58 \\
    400150 & 38.275 & -5.068 & 0.205 &        7 &    --- &      2.33$\times10^{41}$ &   6.92$\times10^{40}$ & 4.85$\times10^{41}$ &    3.73$\times10^{13}$ &  68 \\
    400151 & 38.106 & -5.021 & 0.189 &        7 &    8.87$\times10^{41}$ &      8.89$\times10^{41}$ &   6.81$\times10^{41}$ & 1.10$\times10^{42}$ &    8.08$\times10^{12}$ &  41 \\
    400152 & 38.102 & -4.236 & 0.252 &        6 &    9.08$\times10^{41}$ &      9.82$\times10^{41}$ &   4.30$\times10^{41}$ & 1.60$\times10^{42}$ &    2.96$\times10^{13}$ &  62 \\
    400153 & 34.904 & -4.496 & 0.151 &        9 &    8.00$\times10^{41}$ &      7.99$\times10^{41}$ &   6.94$\times10^{41}$ & 9.04$\times10^{41}$ &    2.82$\times10^{13}$ &  63 \\
    400155 & 33.486 & -5.046 & 0.237 &        7 &    3.91$\times10^{42}$ &      3.91$\times10^{42}$ &   3.19$\times10^{42}$ & 4.63$\times10^{42}$ &    5.43$\times10^{13}$ &  77 \\
    400156 & 33.212 & -4.598 & 0.156 &        7 &    2.93$\times10^{42}$ &      2.92$\times10^{42}$ &   2.67$\times10^{42}$ & 3.18$\times10^{42}$ &    5.87$\times10^{13}$ &  81 \\
    400160 & 35.376 & -4.809 & 0.152 &        7 &    --- &      6.90$\times10^{40}$ &   2.07$\times10^{40}$ & 1.44$\times10^{41}$ &    1.44$\times10^{14}$ & 109 \\
    400161 & 35.255 & -4.764 & 0.151 &        6 &    4.60$\times10^{41}$ &      4.60$\times10^{41}$ &   3.82$\times10^{41}$ & 5.39$\times10^{41}$ &    4.21$\times10^{13}$ &  72 \\
    400162 & 35.153 & -4.623 & 0.198 &        7 &    2.03$\times10^{42}$ &      2.03$\times10^{42}$ &   1.87$\times10^{42}$ & 2.19$\times10^{42}$ &    7.57$\times10^{13}$ &  87 \\
    400163 & 35.978 & -4.794 & 0.044 &        6 &    --- &      5.31$\times10^{39}$ &   1.58$\times10^{39}$ & 1.11$\times10^{40}$ &    3.37$\times10^{12}$ &  32 \\
    400164 & 37.721 & -4.348 & 0.140 &       11 &    1.30$\times10^{43}$ &      1.30$\times10^{43}$ &   1.25$\times10^{43}$ & 1.35$\times10^{43}$ &    7.74$\times10^{13}$ &  89 \\
    400165 & 37.086 & -4.143 & 0.207 &        7 &    4.78$\times10^{41}$ &      5.53$\times10^{41}$ &   2.15$\times10^{41}$ & 9.52$\times10^{41}$ &    1.86$\times10^{14}$ & 116 \\
    400168 & 34.756 & -5.904 & 0.083 &        5 &    9.01$\times10^{40}$ &      9.62$\times10^{40}$ &   4.25$\times10^{40}$ & 1.56$\times10^{41}$ &    6.13$\times10^{12}$ &  39 \\
    400169 & 34.480 & -5.448 & 0.248 &        5 &    7.83$\times10^{41}$ &      7.83$\times10^{41}$ &   5.97$\times10^{41}$ & 9.70$\times10^{41}$ &    1.30$\times10^{12}$ &  22 \\
    400170 & 34.003 & -5.388 & 0.341 &        5 &    6.71$\times10^{42}$ &      6.70$\times10^{42}$ &   5.23$\times10^{42}$ & 8.18$\times10^{42}$ &    1.01$\times10^{14}$ &  91 \\
    400171 & 34.181 & -4.261 & 0.291 &        5 &    --- &      1.59$\times10^{41}$ &   4.76$\times10^{40}$ & 3.30$\times10^{41}$ &    2.01$\times10^{13}$ &  54 \\
    400172 & 34.444 & -4.195 & 0.152 &        5 &    --- &      4.40$\times10^{40}$ &   1.32$\times10^{40}$ & 9.17$\times10^{40}$ &    2.63$\times10^{13}$ &  62 \\
    400173 & 34.924 & -4.009 & 0.203 &        5 &    3.12$\times10^{42}$ &      3.12$\times10^{42}$ &   2.91$\times10^{42}$ & 3.33$\times10^{42}$ &    7.73$\times10^{13}$ &  87 \\
    400174 & 33.892 & -5.774 & 0.288 &        5 &    --- &      4.03$\times10^{41}$ &   1.20$\times10^{41}$ & 8.38$\times10^{41}$ &    8.13$\times10^{13}$ &  86 \\
    400175 & 33.363 & -5.656 & 0.293 &       11 &    --- &      5.43$\times10^{41}$ &   1.63$\times10^{41}$ & 1.14$\times10^{42}$ &    2.84$\times10^{14}$ & 130 \\
    400176 & 33.288 & -5.649 & 0.304 &        8 &    5.51$\times10^{42}$ &      5.51$\times10^{42}$ &   4.37$\times10^{42}$ & 6.66$\times10^{42}$ &    9.07$\times10^{14}$ & 191 \\
    400177 & 33.534 & -5.592 & 0.444 &        5 &    1.94$\times10^{43}$ &      1.94$\times10^{43}$ &   1.79$\times10^{43}$ & 2.09$\times10^{43}$ &    3.30$\times10^{14}$ & 129 \\
    400178 & 33.816 & -4.635 & 0.042 &        5 &    4.78$\times10^{40}$ &      4.78$\times10^{40}$ &   4.08$\times10^{40}$ & 5.48$\times10^{40}$ &    1.42$\times10^{11}$ &  11 \\
    400179 & 33.991 & -4.484 & 0.132 &        6 &    --- &      1.39$\times10^{41}$ &   4.13$\times10^{40}$ & 2.90$\times10^{41}$ &    3.24$\times10^{14}$ & 144 \\
    400180 & 33.148 & -4.279 & 0.196 &        5 &    --- &      1.68$\times10^{41}$ &   5.08$\times10^{40}$ & 3.49$\times10^{41}$ &    5.62$\times10^{13}$ &  79 \\
    400181 & 33.852 & -4.026 & 0.178 &        6 &    1.04$\times10^{42}$ &      1.04$\times10^{42}$ &   7.21$\times10^{41}$ & 1.36$\times10^{42}$ &    2.10$\times10^{13}$ &  57 \\
    400183 & 32.769 & -5.907 & 0.280 &        6 &    3.09$\times10^{42}$ &      3.09$\times10^{42}$ &   2.41$\times10^{42}$ & 3.76$\times10^{42}$ &    2.83$\times10^{13}$ &  61 \\
    400184 & 32.291 & -5.791 & 0.070 &        7 &    --- &      3.27$\times10^{40}$ &   9.82$\times10^{39}$ & 6.81$\times10^{40}$ &    3.45$\times10^{13}$ &  70 \\
    400185 & 32.394 & -5.524 & 0.070 &        5 &    --- &      2.23$\times10^{40}$ &   6.77$\times10^{39}$ & 4.63$\times10^{40}$ &    2.15$\times10^{13}$ &  59 \\
    400186 & 32.342 & -5.237 & 0.072 &        5 &    --- &      2.26$\times10^{40}$ &   6.73$\times10^{39}$ & 4.72$\times10^{40}$ &    1.52$\times10^{12}$ &  25 \\
    400187 & 32.536 & -4.974 & 0.139 &        8 &    2.39$\times10^{42}$ &      2.39$\times10^{42}$ &   2.13$\times10^{42}$ & 2.66$\times10^{42}$ &    8.17$\times10^{13}$ &  91 \\
    400188 & 32.503 & -4.786 & 0.134 &        5 &    --- &      2.42$\times10^{41}$ &   7.26$\times10^{40}$ & 5.04$\times10^{41}$ &    6.64$\times10^{14}$ & 183 \\
    400189 & 32.662 & -4.691 & 0.134 &        5 &    1.92$\times10^{40}$ &      9.36$\times10^{40}$ &   2.88$\times10^{40}$ & 1.91$\times10^{41}$ &    2.66$\times10^{13}$ &  62 \\
    400192 & 31.561 & -5.734 & 0.195 &        5 &    5.45$\times10^{41}$ &      5.62$\times10^{41}$ &   2.76$\times10^{41}$ & 8.64$\times10^{41}$ &    1.51$\times10^{13}$ &  51 \\
    400193 & 32.071 & -5.184 & 0.135 &        6 &    3.59$\times10^{40}$ &      3.59$\times10^{40}$ &   2.88$\times10^{40}$ & 4.31$\times10^{40}$ &    1.25$\times10^{13}$ &  49 \\
    400194 & 31.179 & -4.888 & 0.160 &        7 &    --- &      1.30$\times10^{41}$ &   3.89$\times10^{40}$ & 2.70$\times10^{41}$ &    1.28$\times10^{13}$ &  48 \\
    400195 & 31.581 & -4.445 & 0.238 &        5 &    5.33$\times10^{41}$ &      6.48$\times10^{41}$ &   2.44$\times10^{41}$ & 1.14$\times10^{42}$ &    4.04$\times10^{13}$ &  69 \\
    400196 & 31.262 & -4.289 & 0.240 &        5 &    8.06$\times10^{40}$ &      3.03$\times10^{41}$ &   9.33$\times10^{40}$ & 6.14$\times10^{41}$ &    1.95$\times10^{13}$ &  54 \\
    400197 & 31.740 & -4.302 & 0.205 &        6 &    1.81$\times10^{42}$ &      1.81$\times10^{42}$ &   1.23$\times10^{42}$ & 2.40$\times10^{42}$ &    5.73$\times10^{14}$ & 170 \\
    400199 & 30.933 & -5.997 & 0.130 &        5 &    3.83$\times10^{41}$ &      3.83$\times10^{41}$ &   2.25$\times10^{41}$ & 5.44$\times10^{41}$ &    2.06$\times10^{13}$ &  57 \\
    400200 & 30.632 & -5.946 & 0.214 &        5 &    1.75$\times10^{41}$ &      4.80$\times10^{41}$ &   1.52$\times10^{41}$ & 9.60$\times10^{41}$ &    3.60$\times10^{13}$ &  67 \\
    400201 & 30.603 & -5.910 & 0.194 &        7 &    7.85$\times10^{41}$ &      8.12$\times10^{41}$ &   3.91$\times10^{41}$ & 1.26$\times10^{42}$ &    7.92$\times10^{13}$ &  88 \\
    400203 & 30.524 & -5.262 & 0.189 &        5 &    --- &      1.25$\times10^{41}$ &   3.77$\times10^{40}$ & 2.62$\times10^{41}$ &    2.67$\times10^{13}$ &  61 \\
    400204 & 31.091 & -5.140 & 0.331 &        5 &    4.65$\times10^{41}$ &      8.18$\times10^{41}$ &   2.71$\times10^{41}$ & 1.58$\times10^{42}$ &    6.81$\times10^{14}$ & 172 \\
    400205 & 30.314 & -5.105 & 0.231 &        6 &    1.66$\times10^{43}$ &      1.66$\times10^{43}$ &   1.36$\times10^{43}$ & 1.95$\times10^{43}$ &    6.83$\times10^{13}$ &  83 \\
    400206 & 30.601 & -5.039 & 0.232 &        5 &    7.67$\times10^{41}$ &      7.75$\times10^{41}$ &   4.25$\times10^{41}$ & 1.14$\times10^{42}$ &    5.23$\times10^{13}$ &  76 \\
    400207 & 30.430 & -5.011 & 0.197 &        7 &    1.01$\times10^{42}$ &      1.01$\times10^{42}$ &   5.92$\times10^{41}$ & 1.44$\times10^{42}$ &    3.05$\times10^{13}$ &  64 \\
    400208 & 30.676 & -5.008 & 0.197 &        6 &    8.39$\times10^{41}$ &      8.38$\times10^{41}$ &   6.20$\times10^{41}$ & 1.06$\times10^{42}$ &    3.30$\times10^{13}$ &  66 \\
    400212 & 35.509 & -4.975 & 0.139 &        5 &    --- &      3.42$\times10^{40}$ &   1.02$\times10^{40}$ & 7.08$\times10^{40}$ &    7.44$\times10^{12}$ &  41 \\
    400213 & 35.025 & -4.415 & 0.068 &        5 &    1.39$\times10^{41}$ &      1.39$\times10^{41}$ &   1.31$\times10^{41}$ & 1.46$\times10^{41}$ &    3.08$\times10^{12}$ &  31 \\
    400214 & 35.459 & -4.149 & 0.182 &        5 &    5.13$\times10^{41}$ &      5.13$\times10^{41}$ &   4.18$\times10^{41}$ & 6.10$\times10^{41}$ &    1.07$\times10^{13}$ &  45 \\
    400215 & 35.068 & -4.078 & 0.139 &        5 &    4.30$\times10^{41}$ &      4.30$\times10^{41}$ &   3.69$\times10^{41}$ & 4.93$\times10^{41}$ &    5.76$\times10^{12}$ &  37 \\
    400217 & 36.187 & -5.315 & 0.227 &        5 &    2.61$\times10^{41}$ &      2.65$\times10^{41}$ &   1.39$\times10^{41}$ & 3.95$\times10^{41}$ &    2.20$\times10^{13}$ &  57 \\
    400218 & 36.125 & -5.005 & 0.292 &        5 &    7.33$\times10^{41}$ &      7.38$\times10^{41}$ &   4.32$\times10^{41}$ & 1.05$\times10^{42}$ &    2.50$\times10^{13}$ &  58 \\
    400219 & 36.332 & -4.317 & 0.143 &        5 &    5.89$\times10^{40}$ &      9.00$\times10^{40}$ &   3.04$\times10^{40}$ & 1.69$\times10^{41}$ &    7.27$\times10^{13}$ &  87 \\
    400220 & 36.715 & -4.238 & 0.208 &        5 &    1.66$\times10^{41}$ &      1.72$\times10^{41}$ &   8.17$\times10^{40}$ & 2.70$\times10^{41}$ &    1.85$\times10^{13}$ &  54 \\
    400221 & 36.378 & -4.238 & 0.142 &        9 &    1.50$\times10^{43}$ &      1.50$\times10^{43}$ &   1.47$\times10^{43}$ & 1.53$\times10^{43}$ &    1.14$\times10^{14}$ & 101 \\
    400223 & 37.223 & -5.942 & 0.299 &        6 &    3.13$\times10^{42}$ &      3.14$\times10^{42}$ &   2.07$\times10^{42}$ & 4.21$\times10^{42}$ &    7.01$\times10^{13}$ &  82 \\
    400224 & 37.546 & -5.905 & 0.175 &        5 &    6.65$\times10^{41}$ &      6.68$\times10^{41}$ &   3.76$\times10^{41}$ & 9.67$\times10^{41}$ &    2.80$\times10^{13}$ &  63 \\
    400225 & 37.124 & -5.830 & 0.298 &        5 &    2.46$\times10^{42}$ &      2.45$\times10^{42}$ &   1.71$\times10^{42}$ & 3.22$\times10^{42}$ &    1.69$\times10^{14}$ & 109 \\
    400226 & 37.812 & -5.409 & 0.144 &        8 &    --- &      1.14$\times10^{41}$ &   3.45$\times10^{40}$ & 2.39$\times10^{41}$ &    3.18$\times10^{13}$ &  66 \\
    400227 & 37.645 & -5.179 & 0.277 &        5 &    --- &      3.64$\times10^{41}$ &   1.09$\times10^{41}$ & 7.58$\times10^{41}$ &    1.23$\times10^{14}$ &  99 \\
    400228 & 37.236 & -5.090 & 0.232 &        5 &    3.32$\times10^{41}$ &      3.80$\times10^{41}$ &   1.51$\times10^{41}$ & 6.49$\times10^{41}$ &    4.01$\times10^{12}$ &  32 \\
    400229 & 37.704 & -4.860 & 0.044 &        5 &    3.44$\times10^{40}$ &      3.44$\times10^{40}$ &   2.10$\times10^{40}$ & 4.81$\times10^{40}$ &    7.47$\times10^{11}$ &  20 \\
    400230 & 37.620 & -4.797 & 0.265 &        5 &    2.20$\times10^{42}$ &      2.21$\times10^{42}$ &   1.39$\times10^{42}$ & 3.03$\times10^{42}$ &    3.58$\times10^{14}$ & 142 \\
    400231 & 37.116 & -4.435 & 0.431 &        5 &    1.89$\times10^{43}$ &      1.89$\times10^{43}$ &   1.74$\times10^{43}$ & 2.04$\times10^{43}$ &    2.02$\times10^{14}$ & 110 \\
    400232 & 37.821 & -4.465 & 0.139 &        6 &    1.50$\times10^{42}$ &      1.50$\times10^{42}$ &   1.34$\times10^{42}$ & 1.67$\times10^{42}$ &    2.16$\times10^{13}$ &  58 \\
    400233 & 37.783 & -4.381 & 0.205 &        5 &    --- &      2.64$\times10^{41}$ &   7.89$\times10^{40}$ & 5.50$\times10^{41}$ &    3.61$\times10^{13}$ &  68 \\
    400234 & 38.627 & -5.980 & 0.185 &        6 &    5.86$\times10^{41}$ &      7.53$\times10^{41}$ &   2.71$\times10^{41}$ & 1.36$\times10^{42}$ &    9.53$\times10^{13}$ &  94 \\
    400235 & 38.431 & -5.579 & 0.139 &        8 &    8.51$\times10^{41}$ &      8.51$\times10^{41}$ &   5.39$\times10^{41}$ & 1.17$\times10^{42}$ &    4.37$\times10^{13}$ &  74 \\
    400236 & 37.939 & -4.997 & 0.278 &        6 &    8.04$\times10^{42}$ &      8.04$\times10^{42}$ &   7.15$\times10^{42}$ & 8.92$\times10^{42}$ &    1.43$\times10^{14}$ & 104 \\
    400237 & 38.080 & -4.814 & 0.203 &        5 &    1.61$\times10^{42}$ &      1.61$\times10^{42}$ &   1.32$\times10^{42}$ & 1.90$\times10^{42}$ &    1.17$\times10^{13}$ &  46 \\
    400238 & 38.143 & -4.761 & 0.188 &        5 &    4.00$\times10^{42}$ &      4.00$\times10^{42}$ &   3.59$\times10^{42}$ & 4.42$\times10^{42}$ &    1.09$\times10^{13}$ &  46 \\
    400239 & 38.352 & -4.745 & 0.179 &        5 &    9.00$\times10^{41}$ &      9.01$\times10^{41}$ &   6.41$\times10^{41}$ & 1.16$\times10^{42}$ &    6.09$\times10^{13}$ &  81 \\
    400240 & 38.310 & -4.478 & 0.183 &        6 &    --- &      1.45$\times10^{41}$ &   4.33$\times10^{40}$ & 3.01$\times10^{41}$ &    9.28$\times10^{13}$ &  93 \\
    400242 & 33.169 & -5.515 & 0.043 &        5 &    --- &      1.99$\times10^{40}$ &   5.97$\times10^{39}$ & 4.15$\times10^{40}$ &    3.21$\times10^{13}$ &  68 \\
    400244 & 33.021 & -5.209 & 0.041 &        5 &    1.05$\times10^{41}$ &      1.05$\times10^{41}$ &   7.98$\times10^{40}$ & 1.29$\times10^{41}$ &    1.20$\times10^{12}$ &  23 \\
    400245 & 32.195 & -4.798 & 0.137 &        5 &    7.00$\times10^{41}$ &      7.01$\times10^{41}$ &   5.22$\times10^{41}$ & 8.79$\times10^{41}$ &    8.24$\times10^{13}$ &  91 \\
    400249 & 32.005 & -5.998 & 0.070 &        5 &    6.97$\times10^{40}$ &      6.98$\times10^{40}$ &   4.49$\times10^{40}$ & 9.48$\times10^{40}$ &    9.09$\times10^{11}$ &  21 \\
    400250 & 31.995 & -5.357 & 0.297 &        5 &    5.62$\times10^{42}$ &      5.62$\times10^{42}$ &   4.19$\times10^{42}$ & 7.06$\times10^{42}$ &    1.71$\times10^{14}$ & 110 \\
    400251 & 31.968 & -4.068 & 0.042 &        5 &    3.19$\times10^{40}$ &      3.66$\times10^{40}$ &   1.44$\times10^{40}$ & 6.28$\times10^{40}$ &    1.02$\times10^{13}$ &  47 \\
    400252 & 31.501 & -4.036 & 0.137 &        5 &    --- &      1.13$\times10^{41}$ &   3.39$\times10^{40}$ & 2.35$\times10^{41}$ &    1.35$\times10^{13}$ &  50 \\
    400253 & 30.986 & -5.876 & 0.056 &        5 &    --- &      1.47$\times10^{40}$ &   4.38$\times10^{39}$ & 3.07$\times10^{40}$ &    2.59$\times10^{12}$ &  29 \\
    400258 & 36.802 & -4.600 & 0.297 &        5 &    2.27$\times10^{42}$ &      2.27$\times10^{42}$ &   1.86$\times10^{42}$ & 2.68$\times10^{42}$ &    1.97$\times10^{14}$ & 115 \\
    400260 & 37.278 & -5.567 & 0.030 &        6 &    2.98$\times10^{40}$ &      3.03$\times10^{40}$ &   1.53$\times10^{40}$ & 4.63$\times10^{40}$ &    8.51$\times10^{12}$ &  44 \\
    400261 & 37.834 & -4.801 & 0.079 &        5 &    --- &      2.62$\times10^{40}$ &   7.85$\times10^{39}$ & 5.46$\times10^{40}$ &    9.27$\times10^{12}$ &  45 \\
    400262 & 37.999 & -5.737 & 0.402 &        5 &    7.41$\times10^{42}$ &      7.41$\times10^{42}$ &   5.75$\times10^{42}$ & 9.06$\times10^{42}$ &    9.84$\times10^{13}$ &  88 \\
    400263 & 38.278 & -5.528 & 0.298 &        5 &    4.06$\times10^{42}$ &      4.06$\times10^{42}$ &   3.26$\times10^{42}$ & 4.86$\times10^{42}$ &    2.93$\times10^{14}$ & 131 \\
    400271 & 34.683 & -5.637 & 0.380 &        7 &    --- &      6.54$\times10^{41}$ &   1.96$\times10^{41}$ & 1.37$\times10^{42}$ &    5.05$\times10^{14}$ & 153 \\
    400275 & 34.311 & -4.920 & 0.097 &        5 &    --- &      2.12$\times10^{40}$ &   6.40$\times10^{39}$ & 4.39$\times10^{40}$ &    1.33$\times10^{13}$ &  50 \\
    400282 & 34.147 & -4.325 & 0.154 &        8 &    --- &      6.36$\times10^{40}$ &   1.91$\times10^{40}$ & 1.32$\times10^{41}$ &    4.85$\times10^{13}$ &  76 \\
    400294 & 33.541 & -5.005 & 0.450 &        5 &    9.12$\times10^{42}$ &      9.12$\times10^{42}$ &   6.40$\times10^{42}$ & 1.18$\times10^{43}$ &    1.53E+15 & 215 \\
    400298 & 33.938 & -4.737 & 0.352 &        5 &    --- &      7.53$\times10^{41}$ &   2.26$\times10^{41}$ & 1.57$\times10^{42}$ &    1.57$\times10^{14}$ & 105 \\
    400300 & 33.301 & -4.588 & 0.157 &        5 &    --- &      6.98$\times10^{40}$ &   2.07$\times10^{40}$ & 1.45$\times10^{41}$ &    9.76$\times10^{13}$ &  96 \\
    400305 & 32.480 & -5.652 & 0.237 &        6 &    1.08$\times10^{42}$ &      1.16$\times10^{42}$ &   5.06$\times10^{41}$ & 1.89$\times10^{42}$ &    2.78$\times10^{13}$ &  61 \\
    400308 & 32.548 & -5.316 & 0.137 &        5 &    --- &      8.21$\times10^{40}$ &   2.47$\times10^{40}$ & 1.72$\times10^{41}$ &    9.81$\times10^{12}$ &  45 \\
    400310 & 32.680 & -5.105 & 0.216 &        5 &    1.66$\times10^{42}$ &      1.66$\times10^{42}$ &   1.24$\times10^{42}$ & 2.07$\times10^{42}$ &    2.87$\times10^{13}$ &  62 \\
    400325 & 31.824 & -5.361 & 0.130 &        6 &    4.55$\times10^{41}$ &      4.55$\times10^{41}$ &   3.21$\times10^{41}$ & 5.89$\times10^{41}$ &    3.10$\times10^{13}$ &  66 \\
    400329 & 31.269 & -4.803 & 0.137 &        5 &    --- &      8.59$\times10^{40}$ &   2.56$\times10^{40}$ & 1.79$\times10^{41}$ &    2.79$\times10^{13}$ &  63 \\
    400334 & 31.447 & -4.540 & 0.138 &        5 &    --- &      7.98$\times10^{40}$ &   2.38$\times10^{40}$ & 1.67$\times10^{41}$ &    1.56$\times10^{13}$ &  52 \\
    400360 & 35.059 & -5.388 & 0.277 &        7 &    --- &      2.45$\times10^{41}$ &   7.39$\times10^{40}$ & 5.10$\times10^{41}$ &    2.71$\times10^{14}$ & 129 \\
    400361 & 35.450 & -5.233 & 0.277 &        6 &    5.35$\times10^{41}$ &      5.35$\times10^{41}$ &   3.29$\times10^{41}$ & 7.43$\times10^{41}$ &    4.31$\times10^{13}$ &  70 \\
    400365 & 35.672 & -4.988 & 0.149 &        5 &    --- &      4.47$\times10^{40}$ &   1.34$\times10^{40}$ & 9.35$\times10^{40}$ &    4.15$\times10^{13}$ &  72 \\
    400367 & 35.477 & -4.563 & 0.324 &        8 &    1.44$\times10^{43}$ &      1.44$\times10^{43}$ &   1.38$\times10^{43}$ & 1.51$\times10^{43}$ &    5.92$\times10^{14}$ & 165 \\
    400371 & 35.595 & -4.262 & 0.255 &        6 &    6.54$\times10^{41}$ &      6.52$\times10^{41}$ &   4.44$\times10^{41}$ & 8.61$\times10^{41}$ &    1.33$\times10^{13}$ &  48 \\
    400377 & 36.511 & -5.987 & 0.294 &        7 &    7.95$\times10^{42}$ &      7.94$\times10^{42}$ &   6.43$\times10^{42}$ & 9.46$\times10^{42}$ &    1.11$\times10^{14}$ &  95 \\
    400381 & 36.778 & -5.794 & 0.140 &        5 &    4.51$\times10^{40}$ &      1.36$\times10^{41}$ &   4.27$\times10^{40}$ & 2.70$\times10^{41}$ &    7.65$\times10^{13}$ &  89 \\
    400385 & 36.825 & -5.102 & 0.150 &        5 &    --- &      5.46$\times10^{40}$ &   1.65$\times10^{40}$ & 1.14$\times10^{41}$ &    3.19$\times10^{13}$ &  66 \\
    400386 & 36.019 & -5.016 & 0.320 &        8 &    2.88$\times10^{42}$ &      2.87$\times10^{42}$ &   2.46$\times10^{42}$ & 3.29$\times10^{42}$ &    1.88$\times10^{14}$ & 112 \\
    400389 & 36.544 & -4.897 & 0.085 &        5 &    --- &      2.21$\times10^{40}$ &   6.66$\times10^{39}$ & 4.61$\times10^{40}$ &    2.19$\times10^{13}$ &  59 \\
    400393 & 36.765 & -4.141 & 0.219 &        5 &    --- &      8.60$\times10^{40}$ &   2.58$\times10^{40}$ & 1.78$\times10^{41}$ &    1.96$\times10^{13}$ &  55 \\
    400399 & 37.495 & -5.574 & 0.279 &        6 &    --- &      4.16$\times10^{41}$ &   1.27$\times10^{41}$ & 8.66$\times10^{41}$ &    3.14$\times10^{13}$ &  63 \\
    400401 & 37.333 & -5.369 & 0.290 &        7 &    1.55$\times10^{42}$ &      1.56$\times10^{42}$ &   9.06$\times10^{41}$ & 2.24$\times10^{42}$ &    2.38$\times10^{14}$ & 123 \\
    400414 & 38.137 & -5.568 & 0.216 &        5 &    --- &      1.21$\times10^{41}$ &   3.61$\times10^{40}$ & 2.53$\times10^{41}$ &    1.97$\times10^{12}$ &  26 \\
    400423 & 38.231 & -4.697 & 0.334 &        6 &    1.17$\times10^{42}$ &      1.27$\times10^{42}$ &   5.46$\times10^{41}$ & 2.08$\times10^{42}$ &    6.67$\times10^{13}$ &  79 \\
    400424 & 38.108 & -4.588 & 0.185 &        5 &    --- &      1.48$\times10^{41}$ &   4.43$\times10^{40}$ & 3.10$\times10^{41}$ &    6.26$\times10^{13}$ &  82 \\
    400430 & 33.157 & -5.839 & 0.043 &        5 &    1.11$\times10^{41}$ &      1.11$\times10^{41}$ &   9.57$\times10^{40}$ & 1.26$\times10^{41}$ &    1.65$\times10^{12}$ &  25 \\
    400432 & 33.499 & -5.093 & 0.236 &        6 &    3.74$\times10^{42}$ &      3.74$\times10^{42}$ &   3.11$\times10^{42}$ & 4.37$\times10^{42}$ &    1.16$\times10^{14}$ &  99 \\
    400443 & 30.592 & -5.785 & 0.193 &        6 &    --- &      2.23$\times10^{41}$ &   6.67$\times10^{40}$ & 4.67$\times10^{41}$ &    1.67$\times10^{13}$ &  52 \\
    400459 & 36.798 & -4.392 & 0.176 &        5 &    --- &      5.29$\times10^{40}$ &   1.58$\times10^{40}$ & 1.10$\times10^{41}$ &    1.89$\times10^{12}$ &  26 \\
    400463 & 37.094 & -4.049 & 0.333 &        8 &    3.54$\times10^{41}$ &      1.14$\times10^{42}$ &   3.57$\times10^{41}$ & 2.28$\times10^{42}$ &    5.37$\times10^{14}$ & 159 \\
    400672 & 30.443 & -4.629 & 0.179 &        5 &    4.11$\times10^{40}$ &      2.68$\times10^{41}$ &   8.17$\times10^{40}$ & 5.48$\times10^{41}$ &    7.06$\times10^{13}$ &  85 \\
    400762 & 36.999 & -5.604 & 0.318 &        5 &    2.88$\times10^{42}$ &      2.88$\times10^{42}$ &   2.09$\times10^{42}$ & 3.68$\times10^{42}$ &    6.12$\times10^{13}$ &  77 \\* \bottomrule

\end{longtable}


\bsp	
\label{lastpage}
\end{document}